\documentclass{jfm}
\usepackage{graphicx}
\usepackage{natbib}
\usepackage{xfrac}
\usepackage{comment}
\usepackage{siunitx}
\usepackage[a-1b]{pdfx} 

\usepackage{amsmath}
\allowdisplaybreaks
\usepackage{subcaption}
\usepackage{caption}
\usepackage{xcolor}
\usepackage{siunitx}

\let\realverbatim=\verbatim
\let\realendverbatim=\endverbatim
\renewcommand\verbatim{\par\addvspace{6pt plus 2pt minus 1pt}\realverbatim}
\renewcommand\endverbatim{\realendverbatim\addvspace{6pt plus 2pt minus 1pt}}
\makeatletter
\newcommand\verbsize{\@setfontsize\verbsize{10}\@xiipt}
\renewcommand\verbatim@font{\verbsize\normalfont\ttfamily}
\makeatother

\ifCUPmtlplainloaded \else
  \checkfont{eurm10}
  \iffontfound
    \IfFileExists{upmath.sty}
      {\typeout{^^JFound AMS Euler Roman fonts on the system,
                   using the 'upmath' package.^^J}%
       \usepackage{upmath}}
      {\typeout{^^JFound AMS Euler Roman fonts on the system, but you
                   do not seem to have the}%
       \typeout{'upmath' package installed. jfm.cls can take advantage
                 of these fonts,^^Jif you use 'upmath' package.^^J}%
      }
  \else
  \fi
\fi

\ifCUPmtlplainloaded \else
  \checkfont{msam10}
  \iffontfound
    \IfFileExists{amssymb.sty}
      {\typeout{^^JFound AMS Symbol fonts on the system, using the
                'amssymb' package.^^J}%
       \usepackage{amssymb}%

      }{}
  \fi
\fi

\ifCUPmtlplainloaded \else
  \IfFileExists{amsbsy.sty}
    {\typeout{^^JFound the 'amsbsy' package on the system, using it.^^J}%
     \usepackage{amsbsy}}
    {\providecommand\boldsymbol[1]{\mbox{\boldmath $##1$}}}
\fi

\sbox{\astrutbox}{\rule[-5pt]{0pt}{20pt}}

\title[Journal of Fluid Mechanics]{\LaTeXe\ Input Guide for Authors}
\author{Cheng-Nian Xiao \corresp{\email{chx33@pitt.edu}} \and Inanc Senocak}

\affiliation{Department of Mechanical Engineering and Materials Science, \\ University
of Pittsburgh, 3700 O'Hara St, Pittsburgh, PA 15261, USA }

\date{10 August 2001 and in revised form 10 August 2004}
\pubyear{2001} 
\volume{000}
\pagerange{\pageref{firstpage}--\pageref{lastpage}}
\doi{S002211200100456X}

\begin{document}

\title{ Instabilities of longitudinal vortex rolls in katabatic Prandtl slope flows}
\maketitle
\begin{abstract}

Stationary counter-rotating longitudinal vortex pairs emerge from   one-dimensional Prandtl slope flows under katabatic as well as anabatic conditions due to a linear instability  when the imposed surface heat flux magnitude is sufficiently strong relative to the stable ambient stratification.  For anabatic flows, these vortices have already been identified to exhibit an unique topology that bears a striking resemblance to speaker-wires since they stay coherent as a single unit without the presence of another vortex pair.  Under katabatic conditions and at a constant Prandtl number, we find that the longitudinal vortices emerging at a range of different slope angles  possess the similar topology as their anabatic counterparts. We determine the existence of both fundamental and subharmonic secondary instabilities depending on the slope angle for the most likely transverse base flow wavelength.  Our results indicate that the most dominant instability shifts from a fundamental to subharmonic mode with increasing slope angle.  At shallow slopes, this dynamic contrast with the speaker-wire vortices in anabatic slope flows at the same angle which for which the subharmonic instability is clearly dominant. These   modes are responsible for the bending and movement of single or  multiple speaker-wire vortices, which  may merge or reconnect  to lead to dynamically more unstable states, eventually leading to transition towards turbulence.  We demonstrate that at sufficiently steep slopes, the dynamics of these vortex pairs are dominated by long-wave reconnections or two-dimensional mergers between adjacent pairs.  


\end{abstract}

\section{Introduction}

 Prandtl's model for katabatic and anabatic slope flows is a popular abstraction used to understand the fundamental characteristics of stably stratified flows in cold weather conditions over non-flat terrains, such as nocturnal winds over hills or katabatic flows over the (Ant-)arctic ice sheets \citep{prandtl1942, prandtl1952, gutman1983theory,zardi2013}. The  model can be  solved exactly for laminar conditions to obtain buoyancy and velocity solutions that are sinusoidal profiles damped exponentially with increasing height \citep{shapiro2004,fedorovich2009}. The profiles demonstrate a dominant near-surface along-slope jet which is topped by a weak reverse flow that decays with growing height, as shown in Fig. \ref{fig:prandtlprofile}.
 
 In previous investigations, the  stability of Prandtl's slope flows and travelling waves along a vertical wall has been studied by \cite{gill1969instabilities,mcbain2007instability,maryada2022oblique}.
 We have analyzed the linear stability of  katabatic \citep{xiao2019} and anabatic \citep{xiao2020anabatic} Prandtl slope flows  subject to constant surface heat fluxes  for inclination angles $2^{\circ}<\alpha<70^{\circ}$ and  found the dominating mode to be a stationary longitudinal roll instability at shallow slopes and a traveling wave instability at  slope angles larger than $10^{\circ}$ (anabatic) or $70^{\circ}$(katabatic).
 While the nonlinear travelling wave solutions arising from the corresponding linear mode of the Prandtl slope flows over a vertical wall have been analyzed by \cite{mcbain2007instability, maryada2022oblique}, 
 we instead directed our focus to an investigation of the dynamics of longitudinal rolls arising from the primary longitudinal roll instability under anabatic conditions over a shallow slope with $\alpha=3^{\circ}$ subject to constant surface heat flux \citep{xiao2022speaker}, which is a novel vortex configuration that has not been studied before.  
 Our prior studies have demonstrated that the vortex  instabilities of  stably stratified flows are determined by the slope angle $\alpha$, the  transverse vortex spacing $b_y$, as well as 
  the dimensionless stratification perturbation parameter $\Pi_s$, which can also be interpreted as the surface heat flux normalized by the background stratification. In this present study, we intend to focus on the   dynamics  of stationary vortices in katabatic slope flows.  
 In contrast to anabatic slope flows which appear only fleetingly during morning transition before the positive surface heat flux overpowers the surrounding stable stratification, katabatic conditions can be stable and long-lasting
 as manifested in nocturnal boundary layers or above the (Ant-)arctic ice sheet. Another key difference  is the ability of katabatic conditions  to support stationary longitudinal vortices emerging from a primary instability at very steep slopes up to $65^{\circ}$ \citep{xiao2019}, whereas anabatic conditions do not admit this for slopes larger than $10^{\circ}$\citep{xiao2020anabatic}.

 Numerous experiments under simpler configurations, i.e. without stratification or surface inclination, have confirmed the emergence and persistence of highly organised two-dimensional (2-D) vortex structures in the course of a developing shear
layer downstream of a splitter plate\citep{winant1974vortex,brown1974densityvortices}. These vortices have been noticed to arise under a multitude of different flow
conditions, and appear to be remarkably robust with respect to external disturbances \citep{wygnanski1979perseverance}.  It is evident that any progress in the understanding of turbulent transition will need to  accommodate the dynamics and interaction of these large vortex structures.
In a seminal work on vortex dynamics, \cite{crow1970stability} has studied the linear stability for parallel counter-rotating vortex pairs suspended in   quiescent  air with neutral stratification and discovered that sinusoidal  bending of each individual vortex  arises due to a  symmetric mode now named after him, which then results to vortex merger and  ring formation at precisely those locations where the  the neighboring vortices approach each other due to bending. This work has been extended by \cite{zheng2017importance} to include the analysis of anti-symmetric modes in vortex pairs, whereas \cite{crouch1997instability} studied the dynamics of two unsteadily tumbling vortex pairs.

For more complicated flow configurations which involve the simultaneous presence of  heat transfer, ambient stratification and non-flat surfaces, the  scenario outlined above is expected to be insufficient to describe the entire flow physics.     
However, despite the fact that many real-life phenomena such as geophysical flows contain some or all of these complicating factors,  there has been very few  attempts to properly comprehend vortex instabilities in these challenging situations.

The role of stratification on Stuart vorticies has been 
studied by \cite{miyazaki1992three} who showed that elliptic instabilities causing anti-symmetric distortion of vortex cores are suppressed by the presence of stable stratification. However, \cite{ledizes2009radiative} has found that stable stratification can also destabilize a single vertical columnar vortex  as a result of  resonance with internal gravity waves, which is a so-called radiative instability.
Stable stratification can also give rise to a vortex instability in rotating flows called the "zig-zag"  instability,  first identified and investigated by \citet{billant2000experimental,billant2010zigzag1}. Its formation is  due to additional self induction as well as mutual induction between  vortices as a result of stratification and rotation that are aligned with the main vorticity. Its main effect is  symmetric bending of individual vortices in  co-rotating  pairs and anti-symmetric bending in counter-rotating pairs.

Most of the current and ongoing studies on vortex instabilities such as those listed above have implicitly assumed that  the main merger dynamics depend on no  more than two vortices, which can be either counter-rotating or co-rotating. A notable departure from this tradition is our recent work on the instability of speaker-wire vortices arising in anabatic Prandtl slope flows with surface heating \citet{xiao2022speaker}, which required at least two counter-rotating pairs with four vortices to enable vortex mergers or reconnections. This is due to the fact that all fundamental modes are anti-symmetric such that only  subharmonic instabilities can move  vortices from adjacent pairs closer toward each other.
It has been found that the spacing between adjacent vortex pairs is a main factor for determining the strength of the most dominant modes, which is directly correlated to the likelihood of vortex mergers. An intuitive interpretation of the results presented in \cite{xiao2022speaker} is that the further the initial spacing of the base vortices is from that of the most stable configuration, the stronger is the subharmonic instability of the vortex configuration which would cause vortex mergers or the formation of new vortices in order to attain the most optimal vortex spacing. 
We have also observed in \cite{xiao2022speaker} that  \textit{pure} mergers, i.e. the process of fusing two neighboring vortices into a single one without any bending, can be a dominant dynamics due to the  fact that the two-dimensional vortex instability is almost as strong as the 3D mode with the maximal possible growth rate.\\

The present study aims to extend our previous analysis to investigate the dynamics of the longitudinal rolls that emerge as the  steady configuration of a linear instability under katabatic conditions at shallow slope angles, i.e. subject to surface cooling.
Several distinctions of vortex instabilities  under the Prandtl's model sets our work apart from  the other aforementioned studies on vortex instabilities such as Crow's instability, elliptic instability, zig-zag instability,  and secondary convection rolls. Most significantly,  Prandtl's model includes the following key components missing in other configurations:  Firstly, a constant ambient  stratification that is positioned at an oblique angle to the longitudinal rolls aligned with the main flow direction and secondly,  a solid surface wall containing its own boundary layer as part of the Prandtl base flow.  Unsurprisingly, the combination of these complicating factors  renders a theoretical approach highly difficult and, to the best of our understanding, is not present in the published  literature besides our own earlier related work under anabatic conditions. As demonstrated in \citet{xiao2022speaker}, the dominant vortex instability in anabatic Prandtl slope flows involves two counter-rotating vortex pairs, also termed \textit{speaker-wire vortices}, thus four individual vortices in total.
Following the analysis  in ~\citet{xiao2022speaker}, we apply a modal linear bi-global stability analysis to  identify different  instabilities which can arise from the base flow vortices under katabatic conditions. 
We will further  determine how these instabilities  are controlled by external conditions as well as their role  in the subsequent transition of slope flows to more unstable configurations.

\section{Speaker-wire vortices in katabatic slope flows}

\subsection{Governing equations and characteristic flow scales}
\begin{figure}
 \begin{subfigure}{0.44\textwidth}

	\centering
	\includegraphics[width=0.92\textwidth]{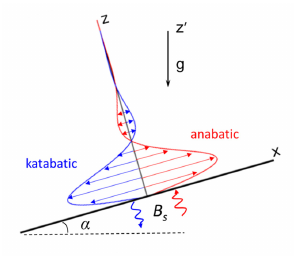}
   \caption{ }
	\label{subfig:sketchslope}
 \end{subfigure}
  \begin{subfigure}{0.54\textwidth}
	\centering
	\includegraphics[width=0.98\textwidth]{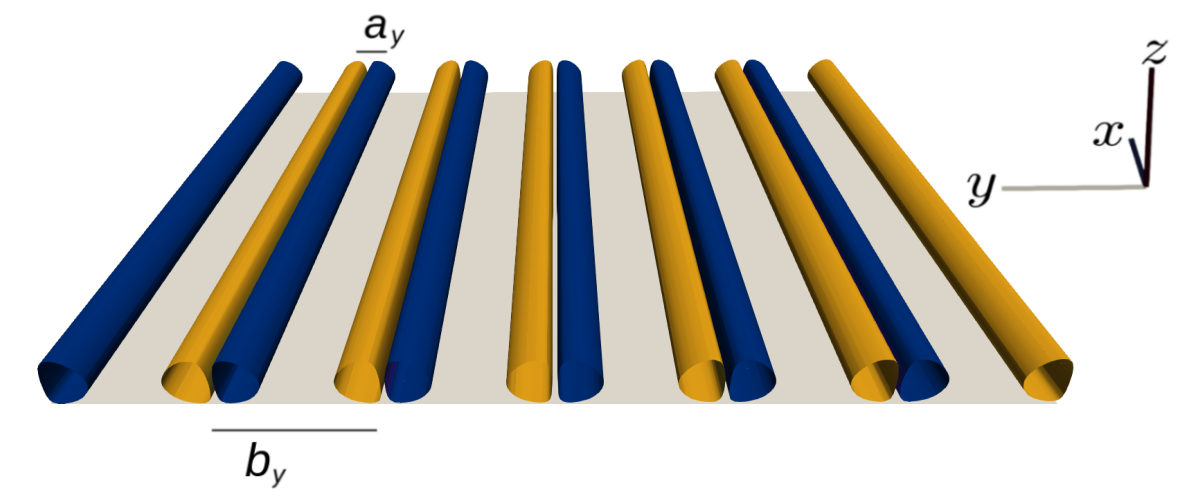}
   \caption{ }
	\label{subfig:slopevortices3d}
 \end{subfigure} \\
 \centering
   \begin{subfigure}{0.62\textwidth}
	\centering
	\includegraphics[width=0.98\textwidth]{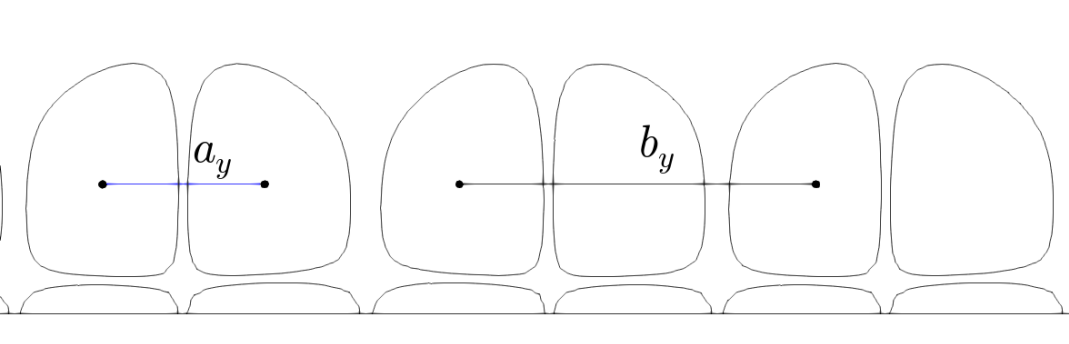}
   \caption{ }
	\label{subfig:slopevortices2d}
 \end{subfigure} 
 
	\caption{ Base flow profiles for slope flows under Prandtl's model: a) Sketch of slope geometry and the rotated coordinate system with 1-D Prandtl slope flow profile; b) Longitudinal vortex pairs (speaker-wire vortices) arising as instability from the 1-D laminar anabatic or katabatic Prandtl slope flow, with definition of coordinate axes,  vortex width $a_y$, and   spacing $b_y$ which is equivalent to the transverse wavelength $\lambda_y$ of the primary longitudinal roll instability; c) cross-slope $yz$ plane with cross-sections of three speaker-wire vortex pairs, indicating of the vortex width  $a_y$ and  vortex spacing $b_y$.  }\label{fig:prandtlprofile}
\end{figure}

Let us consider the idealised Prandtl slope flow configuration as shown in figure \ref{subfig:sketchslope}, where $\alpha$ is the slope angle, and gravity $\mathbf{g}$ acts downward in the vertical direction. A constant uniform negative buoyancy flux $B_S$ is imposed at the surface to achieve katabatic conditions. We consider a rotated Cartesian coordinate system whose $x$ axis is aligned with the planar inclined surface. The direction normal to the slope surface is represented by the $z$ axis, whereas the cross-flow transverse direction is aligned with the $y$ coordinate, as shown in Figs. \ref{subfig:sketchslope}-\ref{subfig:slopevortices3d}. Let $u$ be the along-slope (longitudinal), $v$ be the cross-slope (transverse), and $w$ be the slope-normal velocity components, such that $\mathbf{u}=u_i=[u, v, w]$ is the velocity vector. The unit gravity vector in the rotated coordinate system is then given by  $g_i=(g_1,g_2,g_3)=[\sin\alpha, 0, \cos\alpha]$.

The potential temperature, buoyancy, and the Brunt-V\"ais\"al\"a frequency are denoted by $\theta,b, N$, respectively, where $N$ is related to the ambient potential temperature as $N=\sqrt{\frac{g}{\Theta_r}\frac{\partial \Theta_e}{\partial z'}}$. The buoyancy is defined as a perturbation potential temperature as $b=g   (\Theta-\Theta_e)/ \Theta_r   $, where $\Theta_{r}$ is a reference potential temperature and $\Theta_e$ is the environmental (ambient) potential temperature. The kinematic viscosity and thermal diffusivity of the fluid  are denoted by $\nu,\beta$, respectively, and they are assumed to be constant. 
The transport equations for momentum with a Boussinesq approximation and perturbaton buoyancy fields are written as follows:
\begin{align}
\frac{\partial \mathbf{u} }{\partial t}+\nabla \cdot(\mathbf{u} \otimes \mathbf{u})
=&~ -\frac{1}{\rho}\nabla  p + 
bg\ \mathbf{n}_{\alpha}+ \nu\Delta \mathbf{u} , \label{eqnslopemom}\\
\frac{\partial b}{\partial t} +  \nabla\cdot( b\mathbf{u})  = &~    \beta\Delta b   - N^2 \ (\mathbf{n}_{\alpha} \cdot \mathbf{u}) \label{eqnslopebuoy}.
\end{align}
where $\mathbf{n}_{\alpha}=(\sin\alpha,0,\cos \alpha)$ is the slope-normal unit vector.
The conservation of mass principle  is imposed by a divergence-free velocity field
\begin{align}
\nabla\cdot \mathbf{u}=0.\label{eqnslopecont}
\end{align}

At the surface $z=0$, a negative buoyancy flux $B_{s}$ is imposed to generate anabatic flow conditions against a constant stable ambient stratified environment quantified with $N^2$.   

For the one-dimensional flow problem, \citet{shapiro2004} extended the exact solution of \cite{prandtl1942} to include a constant buoyancy flux  at the surface instead of a constant temperature surface condition and introduced the following characteristic flow scales \cite{fedorovich2009}:
\begin{align}
l_0 = &~ \text{Pr}^{\sfrac{-1}{4}}\ \nu ^{\sfrac{1}{2}} N^{\sfrac{-1}{2}}\sin^{\sfrac{-1}{2}}\alpha , \label{eqnlscale}\\
u_0 = &~ \text{Pr}^{\sfrac{1}{4}}\ \nu ^{\sfrac{-1}{2}}  N^{\sfrac{-3}{2}}B_{s} \sin^{\sfrac{-1}{2}}\alpha , \label{eqnuscale}\\
b_0 = &~ \text{Pr}^{\sfrac{3}{4}}\ \nu ^{\sfrac{-1}{2}}  N^{\sfrac{-1}{2}}B_{s} \sin^{\sfrac{-1}{2}}\alpha ,\label{eqnbscale}
\end{align}
where $\text{Pr}={\nu}/{\beta}$ is the Prandtl number. A time scale $t_0:=l_0/|u_0|=\sqrt{\nu\beta}~N |B_s|^{-1}$, and a shear scale $S_0:=|u_0|/l_0=\sqrt{Pr/\nu}N^{-1}|B_s| $ can also be defined from the above scales. We observe from (\ref{eqnlscale})-(\ref{eqnbscale}) that the length scale characterizing the laminar boundary layer thickness is independent of the surface flux $B_s$, whereas the magnitude of both the reference velocity and buoyancy scale varies linearly with $B_s$. Since only the magnitude of the surface buoyancy flux appears in the flow scale definitions, they are the same for both katabatic and anabatic conditions at the same surface buoyancy flux magnitude but with opposite signs.
Subsequently, these characteristic scales  will be applied to normalize all flow equations and quantities presented herein. 
Specifically, the dimensionless stratification perturbation number  $\Pi_s$ as introduced in \citet{xiao2019},  can be regarded as the imposed surface  buoyancy flux $B_s$ normalised by the background stratification scale $\beta N^2$. This unique parameter is determined from the given external flow parameters as follows: 
\begin{align}
\Pi_s \equiv \frac{|B_s|}{\beta N^2}.
\end{align}
It should be mentioned that the $\Pi_s$ is not restricted to Prandtl slope flows; it is a necessary additional parameter whenever there exists multiple independent stratification mechanisms. In \cite{xiao2022impact}, we demonstrated the significance of $\Pi_s$ as an independent dimensionless parameter in stable  open channels flows that are stratified  by the simultaneous action of surface cooling as well as a static ambient stratification.

\subsection{Steady  speaker-wire vortices as longitudinal rolls }\label{sec_sim2d}
\begin{figure}
\centering

	\centering
	\includegraphics[width=0.6\textwidth]{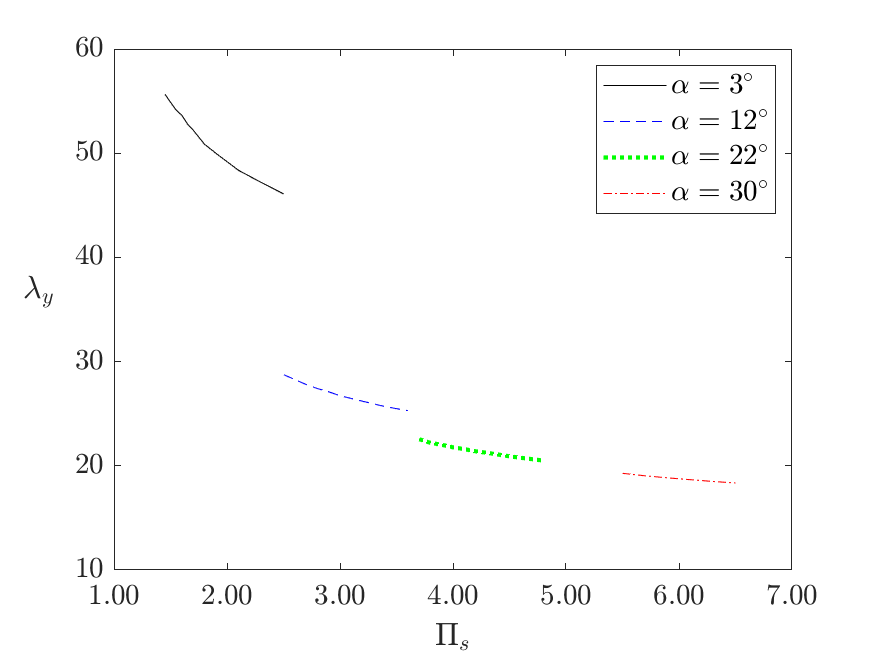}

	\caption{Transverse wavelength of the most dominant  linear instabilities for laminar 1-D katabatic Prandtl slope flows at different slope angles $\alpha$  plotted against $\Pi_s$. }	\label{subfig:fund3_kxcomp}
\end{figure}

The  governing flow equations (\ref{eqnslopemom})-(\ref{eqnslopecont}) can be solved on the 2-D cross-slope $yz$ plane  to arrive at the steady longitudinal rolls that arise as a saturated linear instability for slope angles less than $70^{\circ}$ when $\Pi_s$ is sufficiently large \cite{xiao2020anabatic}. To create the steady vortices that serve as base flows for the secondary stability analysis, the initial flow field is set to be the laminar Prandtl flow profile superposed with a weak sinusoidal disturbance varying along the transverse $y$ direction. For the simulated cross-slope $yz$ plane, the width is chosen to be an integer multiple of the targeted transverse spacing $b_y=\lambda_y$ of the vortex pairs, and the the total height of the  domain is chosen to be at least $50l_0$. Each length scale $l_0$ along both the vertical and transverse direction is resolved by at least two mesh points.   
The evolution of the flow field is simulated until steady state is reached, which happens within a suitable range of $\Pi_s$ values and results in  stationary  vortices that are the saturation of the growing linear instability mode. 
For the spacing of the base vortices, we choose the transverse wavelength $\lambda_y$ of the most dominant longitudinal roll instability, i.e. the primary mode with the largest growth rate for the one-dimensional Prandtl flow profile.
The dependence of $\lambda_y$
as a function of the stratification parameter $\Pi_s$ for different slope angles is shown in Fig. \ref{subfig:fund3_kxcomp}, which indicates a clear decrease of the transverse wavelength with increasing value of $\Pi_s$ and slope angle $\alpha$. Further details about the structure of the vortices that serve as base flow for the secondary stability analysis are provided along with the stability results below.

\section{Linear secondary instability analysis of speaker-wire vortices}

Let $(U,V,W,B)$ be the 2-D flow profile of the steady longitudinal rolls as computed from Eqn. (\ref{eqnslopemom})-(\ref{eqnslopebuoy}),    and assuming that modal disturbances  to this base flow are  of the form \[\mathbf{q}(x,y,z,t) =\left[\hat{u}(y,z), \hat{v}(y,z), \hat{w}(y,z), \hat{p}(y,z),\hat{b}(y,z)\right]\exp({ik_x x + \omega t}),\]
then the equations governing the evolution of the flow disturbances approximated to first the order have the following form:
\begin{align}
ik_x\hat{u}+\frac{\partial \hat{v}}{\partial y}+ \frac{\partial \hat{w}}{\partial z} &= 0,\label{eqnslopelincont}\\
\omega \hat{u}+iUk_x\hat{u} +\frac{\partial U}{\partial y} \hat{v} + \frac{\partial U}{\partial z}\hat{w} & 
+\frac{\partial \hat{u}}{\partial y} V +  \frac{\partial \hat{u}}{\partial z}W \notag \\
&=  -ik_x\hat{p} -\frac{ Pr}{\Pi_s}\sin\alpha\left(-k_x^2\hat{u}+   \frac{\partial^2\hat{u}}{\partial y^2}+\frac{\partial^2\hat{u}}{\partial z^2} +\hat{b}\right),\\ 
\omega \hat{v}+iUk_x\hat{v}+\frac{\partial V}{\partial y}\hat{v}& +\frac{\partial V}{\partial z}\hat{w} 
+\frac{\partial \hat{v}}{\partial y}V+\frac{\partial \hat{v}}{\partial z}W  \notag \\
&= -\frac{\hat{p}}{\partial y}-\frac{Pr}{\Pi_s}\sin\alpha\left(-k_x^2\hat{v}+\frac{\partial^2\hat{v}}{\partial y^2} +\frac{\partial^2\hat{v}}{\partial z^2} \right),\\
\omega \hat{w}+iUk_x\hat{w} +\frac{\partial W}{\partial y}\hat{v}& +\frac{\partial W}{\partial z}\hat{w} +\frac{\partial \hat{w}}{\partial y}V+ \notag\\ \frac{\partial \hat{w}}{\partial z}W 
= -\frac{\partial\hat{p}}{\partial z} & -\frac{ Pr}{\Pi_s}\sin\alpha\left(-k_x^2\hat{w}+\frac{\partial^2\hat{w}}{\partial y^2} +\frac{\partial^2\hat{w}}{\partial z^2}+ \hat{b} \cot\alpha  \right),\\
\omega \hat{b}+iUk_x\hat{b}+\frac{\partial B}{\partial y}\hat{v}& +\frac{\partial B}{\partial z}\hat{w} +\frac{\partial \hat{b}}{\partial y}V+\frac{\partial \hat{b}}{\partial z}W  \notag \\
&= -\frac{\sin\alpha}{\Pi_s}\left(-k_x^2\hat{b}+\frac{\partial^2\hat{b}}{\partial y^2} +\frac{\partial^2\hat{b}}{\partial z^2} -( \hat{u}+\hat{w}\cot\alpha) \right), \label{eqnslopelinlast}
\end{align}
where $\hat{u},\hat{v},\hat{w},\hat{p},\hat{b}$ describe the shape of the flow disturbance along the slope normal and transverse directions,  normalised by the flow scales given in  (\ref{eqnlscale})-(\ref{eqnbscale}). The normalised base flow field describing the steady vortices is denoted by $(U,V,W, B)$, and the slope angles studied here are  $\alpha =3^{\circ},  12^{\circ}, 22^{\circ}, 30^{\circ}$. \\

The linearised equations for bi-global stability analysis as shown above can be written as a generalised eigenvalue problem  in the following way:
\begin{align}
A(k_x)\mathbf{\hat{q}}(y,z)=\omega B(k_x)\mathbf{\hat{q}}(y,z),
\label{eig}
\end{align}
The  shape of the complex disturbance  vector 
\[
\mathbf{\hat{q}}(y,z)=[\hat{u}(y,z),\hat{v}(y,z),\hat{w}(y,z),\hat{p}(y,z),\hat{b}(y,z
)]^T
\]
varies in the slope-normal (z) and transverse direction (y), where $(\hat{u},\hat{v},\hat{w})$ are the along-slope, cross-slope (transverse) and slope-normal disturbance velocity components.
As a bi-global stability analysis, the slope-normal and transverse dimensions are fully resolved and the disturbance variation along the streamwise direction is approximated by only one Fourier mode with wave number $k_x$. When $k_x$ is zero, then the corresponding mode is 2-D without any streamwise variation, whereas a positive $k_x$ implies a full 3-D disturbance.  The appropriate boundary conditions for this problem are no slip for disturbance velocities at $z=0$ and free-slip at $z\rightarrow \infty$, whereas for the buoyancy disturbance, $\partial \hat{b}/\partial z|_{z=0}= \partial \hat{b}/\partial z |_{z\rightarrow \infty}=0$ are imposed. The slope-normal derivative of pressure disturbance $\hat{p}$ is also set to zero at both $z=0$ and $z\rightarrow \infty$.  On both transverse boundaries, periodic conditions are imposed for all variables. The generalised eigenvalue problem (\ref{eig}) is discretised via  spectral elements in the transverse plane,
 which is available in Nektar++ \cite{nektar++}. For a base flow containing two full transverse spatial periods, i.e. two speaker-wire vortex pairs, the cross-slope dimension  $L_y=2\lambda_y$ is chosen for the simulation domain; 2 degrees of freedom are used to resolve a unit length scale $l_0$ in the cross-slope and vertical directions, and the resulting generalised eigenvalue problem are solved with the modified Arnoldi algorithm as implemented in Nektar++. Linear stability of the problem is associated with the real part of the eigenvalues $\omega$, where $\Re\{ \omega\}>0$ represents a positive exponential growth for the corresponding eigenmode, thus an unstable mode. The imaginary part of $\omega$
is the temporal oscillation frequency for the corresponding eigenmode, and $\Im\{ \omega\}=0$ represents a stationary mode.

\subsection{Instabilities of katabatic speaker-wire vortices at different  slope angles}
To investigate the secondary linear instability of speaker-wire vortices,  eigenvalues with the highest maximal real values  for a range of streamwise wave numbers $k_x$ are computed for different   stratification perturbation parameters $\Pi_s$ of the base flow at different slope angles $\alpha$. The transverse wavelength $\lambda_y$ of the   base flow vortices was chosen to be the most dangerous  wavelength at the given $\Pi_s$ as determined from linear stability analysis of the 1D Prandtl flow profile, which is shown in Fig.~\ref{subfig:fund3_kxcomp}. Thus in contrast to the study for vortices under anabatic conditions \citep{xiao2022speaker}, the focus here is on the effects of the slope angle instead of the vortex spacing.  Throughout our analysis, we also assume a constant Prandtl number $Pr =0.71$ for all the cases.

We observe from Fig. \ref{subfig:fund3_kxcomp} that the  slope inclination $\alpha$ exerts a significant effect on the optimal transverse wavelength of the base vortices when all other flow parameters are held constant. Hence, we expect that the secondary instabilities arising from these base flow vortices will also be qualitatively different depending on $\alpha$. At a given slope angle and Prandtl number, three independent parameters determine the growth rates and oscillation frequencies of the secondary instability, which are the stratification parameter $\Pi_s$, the streamwise instability wave number $k_x=2\pi/\lambda_x$ and the transverse base flow wave number $k_y=2\pi/\lambda_y$, where $\lambda_y=b_y$ equals the  spacing of the base speaker-wire vortex. In the following, we will separately present and discuss the results of the stability analysis for base speaker-wire vortex  at four different slope angles $\alpha$, going from $\alpha=3^{\circ}$ to  $\alpha=30^{\circ}$.
The base flow used for  modal analysis of each case consists of two speaker-wire vortices arising from the primary linear instability mode, i.e. the transverse domain size is twice the  wavelength of the primary vortex instability as described in \cite{xiao2020anabatic}.  This choice of domain size  ensures that potential subharmonic eigenmodes with twice the transverse wavelength of the base vortices can be picked up. 
\\

\subsubsection{Shallow slope: $\alpha=3^{\circ}$}

\begin{figure}

 \vspace{15pt}
   
		\centering
	\includegraphics[width=0.7\textwidth]{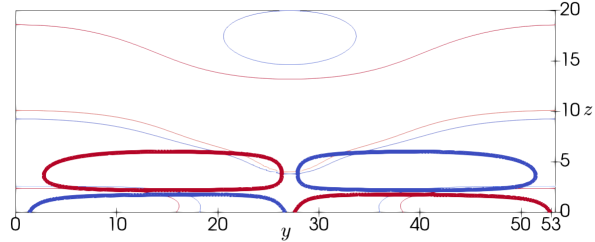}

	\caption{ Visualization  of streamwise vorticity magnitude for the fundamental mode along with base vortices  on cross-slope $yz$ vertical plane  in katabatic  flow  at  $\alpha=3^{\circ}$ and $\Pi_s=1.6$ with streamwise wave number $k_x=0.07$. Thick contour lines represent the base flow, and thin lines are from the instability mode.  Contours are drawn at 8\% of maximal magnitude in both cases. Only half of the total transverse  domain   is shown,  equalling the transverse wavelength $\lambda_y$ of the base vortices.} 	\label{fig:fund03_contour2d}
\end{figure}

At the shallow slope  $\alpha=3^{\circ}$, it turns out that the transverse spatial period of all the secondary instability  modes   equals the transverse wavelength $\lambda_y$ of the base vortices which is also the spacing $b_y$  between the core of adjacent vorte pairs,  hence they are all \textit{fundamental} modes. The shape of this mode on the transverse plane  is visualized via the contour plot for the streamwise vorticity $\omega_x$ as shown in Fig. \ref{fig:fund03_contour2d}. Thus,  this observation is in contrast with the anabatic slope flow at the same angle $\alpha=3^{\circ}$ where the most dominant instability is a subharmonic mode \cite{xiao2022speaker}. 

\begin{figure}
 \begin{subfigure}{0.5\textwidth}

	\centering
	\includegraphics[width=1\textwidth]{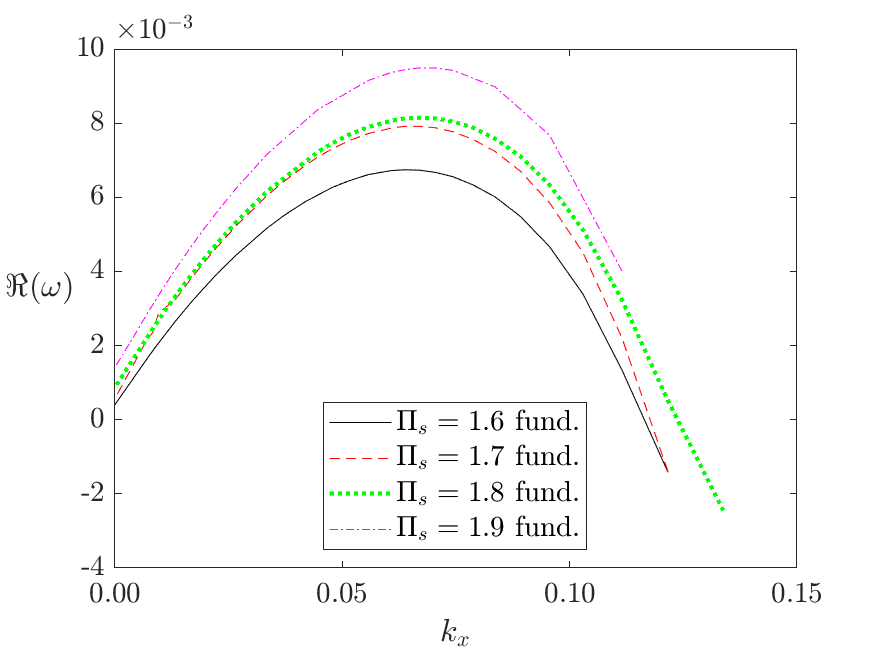}
   \caption{ }
	\label{subfig:fund03_gr}
 \end{subfigure}
  \begin{subfigure}{0.5\textwidth}
	\centering
	\includegraphics[width=1\textwidth]{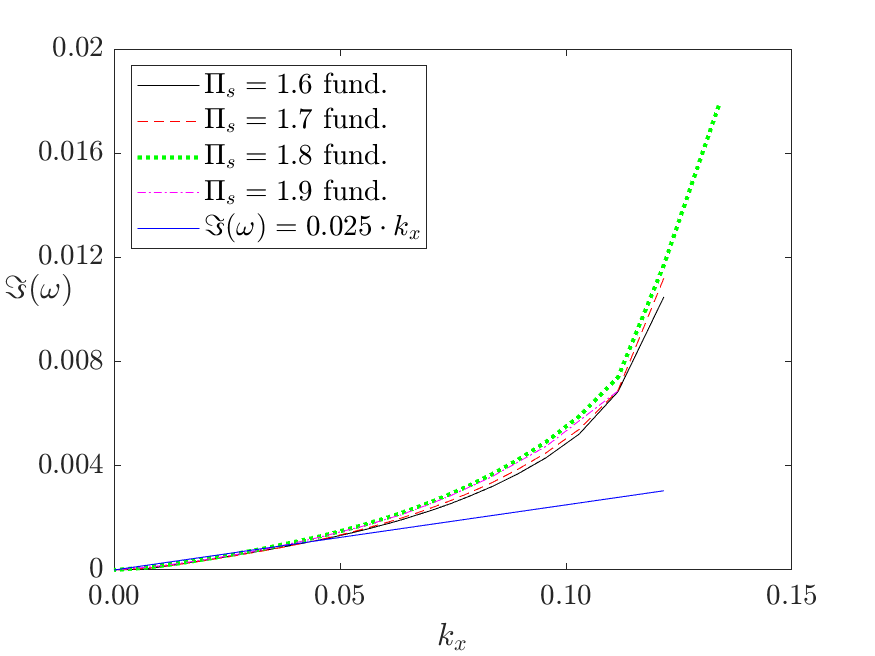}
   \caption{ }
	\label{subfig:fund03_osc}
 \end{subfigure}

	\caption{Fundamental  modes of katabatic speaker-wire vortices at slope angle $\alpha=3^{\circ}$   for different values of $\Pi_s$ as a function of the longitudinal wave number $k_x$: Growth rate $\Re(\omega)$ (a) and  oscillation frequency $\Im(\omega)$ (b). The linear fit for the dispersion relation between oscillation frequency and streamwise wave number is also shown in (b). }
\end{figure}

Figure \ref{subfig:fund03_gr} shows the  growth rates of the most unstable secondary modes as a function of streamwise wavenumbers  $k_x$ within the range $[0,0.15]$ for four different values of stratification parameter $\Pi_s =1.6,1.7,1.8,1.9$. The growth rate of any mode at any streamwise wavenumber $k_x$ grows with increasing $\Pi_s$, which implies a stronger surface cooling for the katabatic slope flow. 
2-D modes  have zero longitudinal wave number, i.e. $k_x=0$, and thus only vary on the transverse $yz$ plane but are constant along the streamwise  direction.

For the 3-D modes, which have non-zero wave numbers $k_x>0$, we observe from figure~\ref{subfig:fund03_gr} that at large wave numbers $k_x>0.15$, the growth rates tend to increase   with decreasing $k_x$ until reaching a maximal value exceeding the 2-D  growth rate ($k_x=0$) at a optimal wave number $k_x\approx 0.07$, and from that point on they decrease with decreasing wavenumber to reach the 2-D growth rate at $k_x=0$. In contrast to both the subharmonic and fundamental eigenmodes of speaker-wire vortices under anabatic conditions at the same slope angle of $\alpha=3^{\circ}$  \citep{xiao2022speaker},  the most  dominant fundamental modes at katabatic conditions are clearly 3-D since the growth rate of the mode at the optimal non-zero wave number is an order of magnitude larger than that of the zero-wavenumber 2-D mode.

All 3-D fundamental modes with positive longitudinal wavenumbers $k_x>0$ are oscillatory whose frequency increases monotonically with growing $k_x$ or decreasing wavelength, as shown in figure \ref{subfig:fund03_osc}. The 2-D mode with $k_x=0$  is  stationary, i.e. with  zero imaginary part of its eigenvalue. At small streamwise wavenumbers $k_x$, the oscillation frequencies of the 3-D mode  decrease to zero with decreasing $k_x$ to converge to the stationary 2-D mode at $k_x=0$. As figure \ref{subfig:fund03_osc} shows, within the small wave number range, the  normalised frequencies of all three cases shown here appear to obey a simple linear dispersion relation given by $\Im(\omega)=\eta\cdot k_x$, where  $\eta \approx 0.025$  is empirically determined to fit all three curves. The accuracy of this fit shows that the group speed  of the long-wave fundamental vortex instabilities given by $c=\frac{\partial \Im(\omega)}{\partial k_x}=\eta$ is nearly constant  and equals $\eta =0.025$. However, this simple relation no longer holds for modes with  wave numbers $k_x>0.05$ whose oscillation frequencies increase faster than linear with growing wave numbers.

Existence of  both 2-D  and 3-D fundamental vortex instabilities has also been identified in the study of co-rotating Stuart vortex arrays  by \cite{pierrehumbert1982vortexinstab} or as  instabilities in a shear layer by \cite{corcos1984mixing}, where they are shown to be responsible for the displacement or bending of neighboring vortices. In contrast to the anabatic slope flow at a slope angle of $\alpha=3^{\circ}$ \citep{xiao2022speaker}, there are no subharmonic modes in the  katabatic slope flow with angle $\alpha=3^{\circ}$ . As will be shown later, the fundamental modes for this case as well as at larger slope angles are all anti-msymmetric. This suggest that at the initial onset of instabilities, the neighboring vortex tubes will bend and move synchronously in parallel fashion. This is confirmed by direct numerical simulations via solving the full nonlinear governing equations, as shown in the animation available as supplementary movie 1. From the animation,  the consistent growth of the anti-symmetric fundamental mode  which eventually generates novel structures on top of the original vortices can be seen.

\subsubsection{ Moderately steep slopes $\alpha=12^{\circ}$}

At a slope angle of $\alpha=12^{\circ}$, both fundamental and subharmonic secondary instabilities are found to emerge. The transverse wavelength of fundamental   mode equals  the spacing $b_y$ between adjacent base flow speaker-wire vortices, whereas the subharmonic mode have transverse wavelengths twice as large. Hence the additional appearance of the subharmonic mode is a consequence of the increasing slope angle compared to the shallow slope at $\alpha=3^{\circ}$.

\begin{figure}

 \vspace{15pt}
 \centering
   \begin{subfigure}{0.65\textwidth}
		
	\includegraphics[width=0.99\textwidth]{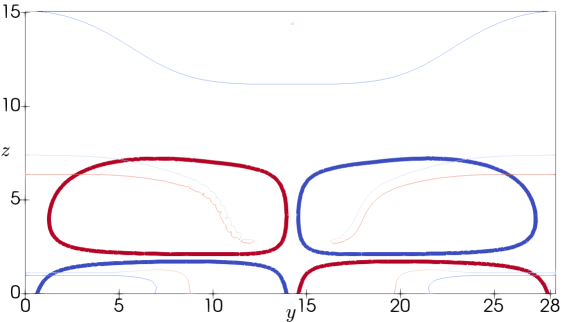}
   	\caption{}
	\label{subfig:fund12_contour2d}
 \end{subfigure}

      \begin{subfigure}{0.75\textwidth}
      	\centering
	\includegraphics[width=0.99\textwidth]{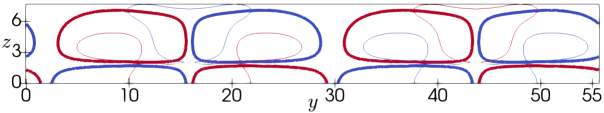}
   	\caption{}
	\label{subfig:subh12_contour2d}
 \end{subfigure}
 
	\caption{ Visualization  of streamwise vorticity magnitude for the fundamental mode on cross-slope $yz$ vertical plane  in katabatic  flow  at   $\alpha=12^{\circ}$ and $\Pi_s=2.6$: (a) fundamental mode with $k_x=0.1$; (b) subharmonic mode with $k_x=0.015$. Thick contour lines represent the base flow, and thin lines are from the instability mode.  Contours are drawn at 8\% of maximal magnitude in both cases.  } 
\end{figure}

 The shapes of the subharmonic and fundamental modes on the transverse plane  are visualized together with the base flow vortices via the contour plot for the streamwise vorticity $\omega_x$ as shown in Figs. \ref{subfig:fund12_contour2d} and \ref{subfig:subh12_contour2d}. It can be seen that compared to the $3^{\circ}$ case as shown in Fig. \ref{fig:fund03_contour2d}, the base vortices are around half as wide along the transverse direction, but also clearly taller in the vertical. As per definition, the subharmonic modes have a transverse wavelength twice as large as their fundamental counterparts.

\begin{figure}
 \begin{subfigure}{0.5\textwidth}
	\includegraphics[width=1\textwidth]{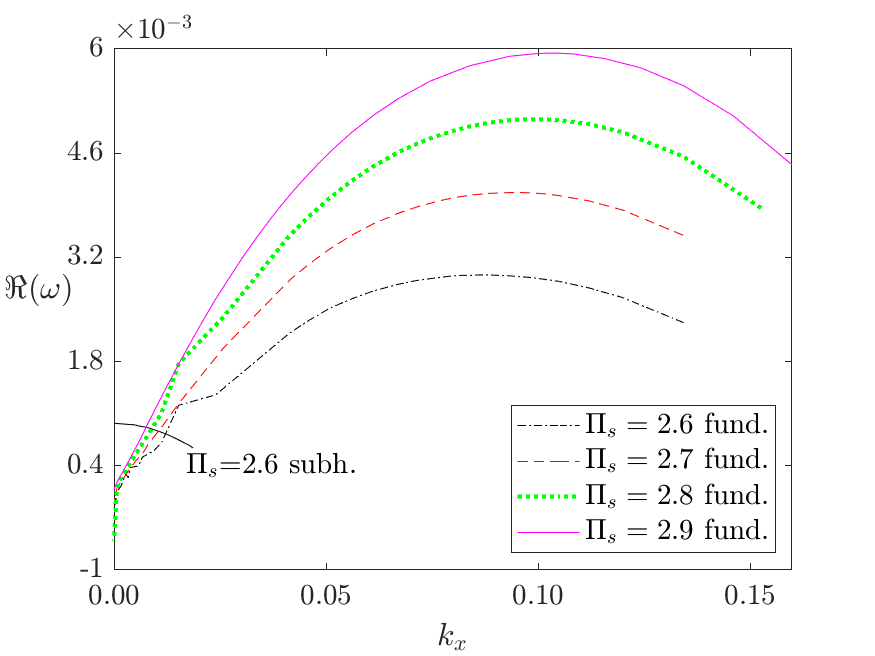}
   \caption{ }
	\label{subfig:fund12_gr}
 \end{subfigure}
  \begin{subfigure}{0.5\textwidth}
	\includegraphics[width=1\textwidth]{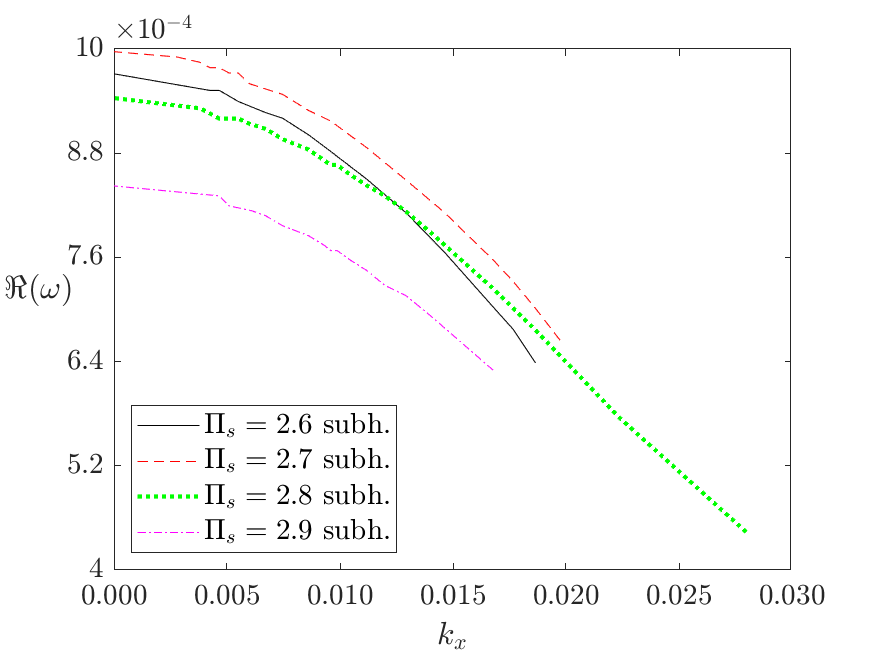}
   \caption{ }
	\label{subfig:subh12_gr}
 \end{subfigure}
 \vfill
 \vspace{15pt}
   \begin{subfigure}{0.5\textwidth}
\centering
	\includegraphics[width=1\textwidth]{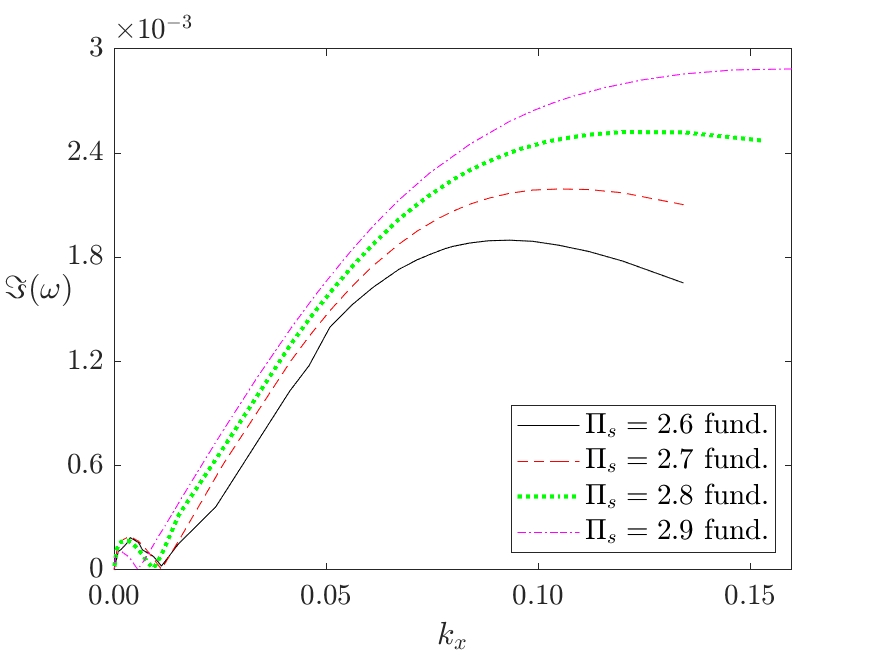}
   	\caption{}
	\label{subfig:fund12_osc}
 \end{subfigure} 
    \begin{subfigure}{0.5\textwidth}
	\includegraphics[width=1\textwidth]{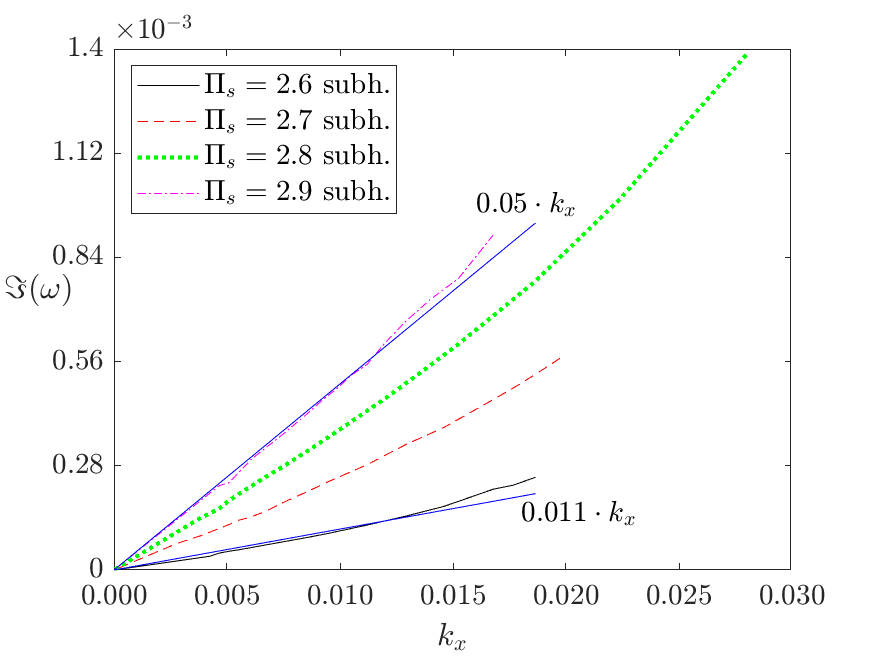}
   	\caption{}
	\label{subfig:subh12_osc}
 \end{subfigure}
 
	\caption{ Comparison between fundamental and subharmonic modes of katabatic speaker-wire vortices at slope angle $\alpha=12^{\circ}$   for different values of $\Pi_s$ as a function of the longitudinal wave number $k_x$  : (a)-(b) Growth rates $\Re(\omega)$;  (c)-(d) oscillation frequency $\Im(\omega)$. Fundamental modes are shown on the left(a,c), and subharmonic modes are shown on the right(b,d). The linear fit for the relations between oscillation frequency  and streamwise wave number of the subharmonic mode is also shown in  (d).  } \label{fig:fund12_gr}
\end{figure}

Figure \ref{subfig:fund12_gr} presents the  growth rates of the most unstable fundamental modes as a function of streamwise wavenumbers  $k_x$ within the range $[0,0.15]$ for four different values of normalized surface heat flux $\Pi_s =2.6,2.7,2.8,2.9$.  
The  growth rates of the strongest subharmonic modes for streamwise wavenumbers  $k_x$ within  $[0,0.03]$ are shown  in Fig.  \ref{subfig:subh12_gr}. The smaller  wavenumber range for the  strongest subharmonic modes displayed here implies that their streamwise wavelengths are longer compared to the fundamental modes. For comparison, the growth rate for the strongest subharmonic mode at $\Pi_s=2.6$ is also displayed together with all the fundamental modes Fig. \ref{subfig:fund12_gr}. 

Compared to the shallow slope with $\alpha=3^{\circ}$, larger $\Pi_s$ values are needed to sustain the base vortices that arise as the primary roll instability from the 1D Prandtl slope flow profile \citep{xiao2019}. It is evident from Fig. \ref{subfig:fund12_gr} that a higher value of $\Pi_s$, implying a larger normalized surface heat flux magnitude, increases the growth rate of all fundamental modes at any given wave number. However, the same $\Pi_s$ dependence is not true for the subharmonic modes. As Fig. \ref{subfig:subh12_gr} shows, the  growth rate for subharmonic modes attains its maximum at $\Pi_s=2.7$ and is smaller for larger as well as smaller $\Pi_s$ values.

From figure \ref{subfig:fund12_gr},  we observe that for  modes with sufficiently large wave numbers $k_x>0.01$, the fundamental mode is stronger than its subharmonic counterpart. The  fundamental modes  attain the strongest growth at an optimal non-zero wave number $k_x$ between 0.08 and 0.1, and they become monotonically weaker with decreasing  wave number below that value. On the other hand, as we observe from figure \ref{subfig:subh12_gr}, the subharmonic modes gain strength with decreasing wave number such that the most dominant mode is purely 2-D at $k_x=0$. Hence, for very large streamwise wavelengths with $k_x<0.01$, fundamental modes are weaker than their subharmonic counterparts.

From figure \ref{subfig:fund12_osc}, we observe that all  fundamental modes with positive streamwise wavenumbers $k_x>0$ are oscillatory.  For $k_x=0$, the fundamental modes are all stationary, and their  frequencies as a function of $k_x$ attains one  local maximum in the low-wavenumber range at  $k_x\approx 0.004$ and begin to increase with growing wavenumbers for $k_x>0.012$, as shown in figure \ref{subfig:fund12_osc}.
However, at higher wavenumbers beyond $k_x=0.1$,  this trend   no longer holds, as the fundamental mode frequencies either stagnate (for $\Pi_s=2.8,2.9$) or even slightly decrease (for $\Pi_s=2.6,2.7$) with increasing $k_x$. On the other hand, figure \ref{subfig:subh12_osc} shows that the frequency of subharmonic modes increases monotonically with growing wavenumber. At $k_x=0$, the 2-D subharmonic mode can be seen to be stationary.  As Fig. \ref{subfig:subh12_osc} shows, for wavenumber values less than $k_x=0.01$, the  normalised frequencies for all four $\Pi_s$ values  fit a linear dispersion relation where the group velocity given by $\eta=\frac{\partial \Im(\omega)}{\partial k_x}$ increases with growing $\Pi_s$   from $\eta=0.011$ at $\Pi_s=2.6$  to $\eta=0.05$ for $\Pi_s=2.9$. This differs from the behavior of subharmonic modes in anabatic slope flows where the group velocity $\eta$ remains the same for different $\Pi_s$ values.
Fig. \ref{subfig:subh12_osc} shows that at larger wavenumbers, the linear dispersion relation no longer holds because the oscillation frequencies  increase at a faster rate than linear rate with growing wavenumbers for $k_x>0.01$.  

The above-mentioned properties of fundamental and subharmonic modes have significant impact on the  dynamics of the base vortices. Since the strongest subharmonic mode is 2-D, we can expect fusion between neighboring vortex pairs instead of wavy reconnections that result in vortex rings like in the Crow instability found in \cite{crow1970stability}. The fundamental modes which dominate at shorter wavelengths are antisymmetric and will bend each individual vortex parallel to its adjacent neighbors. 
This co-existence of both dynamics  can be seen  from animations obtained from direct numerical integration  of the full nonlinear governing equations as shown in the  supplementary movie 2. The animation displays the initial presence of the anti-symmetric fundamental mode which gradually gives way to the subharmonic mode dynamics which moves neighboring vortex pairs closer for mergers.

\subsubsection{ Steep slopes $\alpha=22^{\circ}$}

\begin{figure}

 \vspace{15pt}
 \centering
   \begin{subfigure}{0.5\textwidth}
	\includegraphics[width=0.8\textwidth]{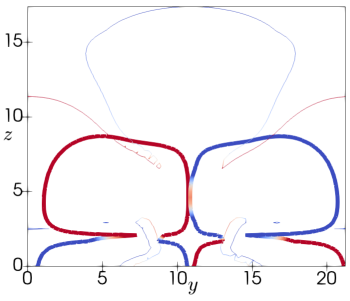}
   	\caption{}
	\label{subfig:fund22_contour2d}
 \end{subfigure}

      \begin{subfigure}{0.7\textwidth}
      
	\includegraphics[width=0.99\textwidth]{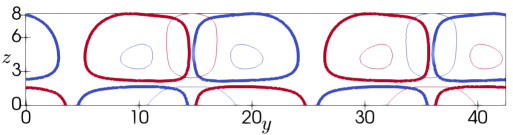}
   	\caption{}
	\label{subfig:subh22_contour2d}
 \end{subfigure}
 
	\caption{ Visualization  of streamwise vorticity magnitude for the fundamental mode on cross-slope $yz$ vertical plane  in katabatic  flow  at    $\alpha=22^{\circ}$ and $\Pi_s=3.9$: (a) fundamental mode with $k_x=0.13$; (b) subharmonic mode with $k_x=0.015$. Subharmonic modes have widths twice as large as fundamental ones. Thick contour lines represent the base flow, and thin lines are from the instability mode.  Contours are drawn at 8\% of maximal magnitude in both cases. 
  } 
\end{figure}

For even steeper  slopes with an angle of $\alpha=22^{\circ}$, subharmonic and fundamental secondary instabilities again  manifest themselves. The typical shapes of these two different kinds of modes on the transverse plane  are visualized  alongside the base flow vortices via the contour plot for the streamwise vorticity $\omega_x$ as shown in Figs. \ref{subfig:fund22_contour2d} and \ref{subfig:subh22_contour2d}.    Compared to the  previous two cases at smaller angles as shown in Fig. \ref{fig:fund03_contour2d} and Figs. \ref{subfig:fund12_contour2d}-\ref{subfig:subh12_contour2d}, the base vortices are narrower along the transverse direction, but also slightly taller in the vertical direction. 

\begin{figure}
 \begin{subfigure}{0.5\textwidth}

	\centering
	\includegraphics[width=1\textwidth]{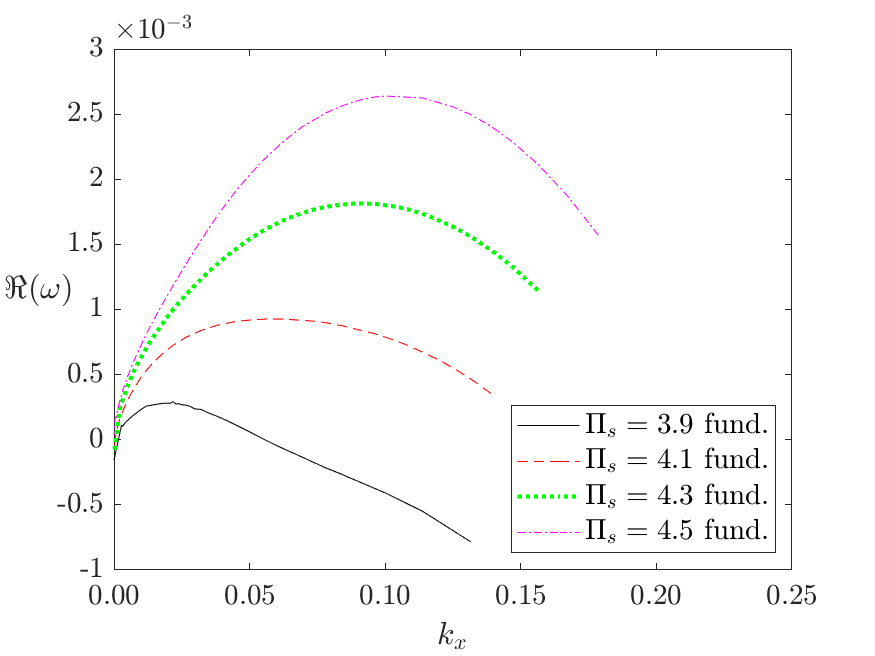}
   \caption{ }
	\label{subfig:fund22_gr}
 \end{subfigure}
  \begin{subfigure}{0.5\textwidth}
	\centering
	\includegraphics[width=1\textwidth]{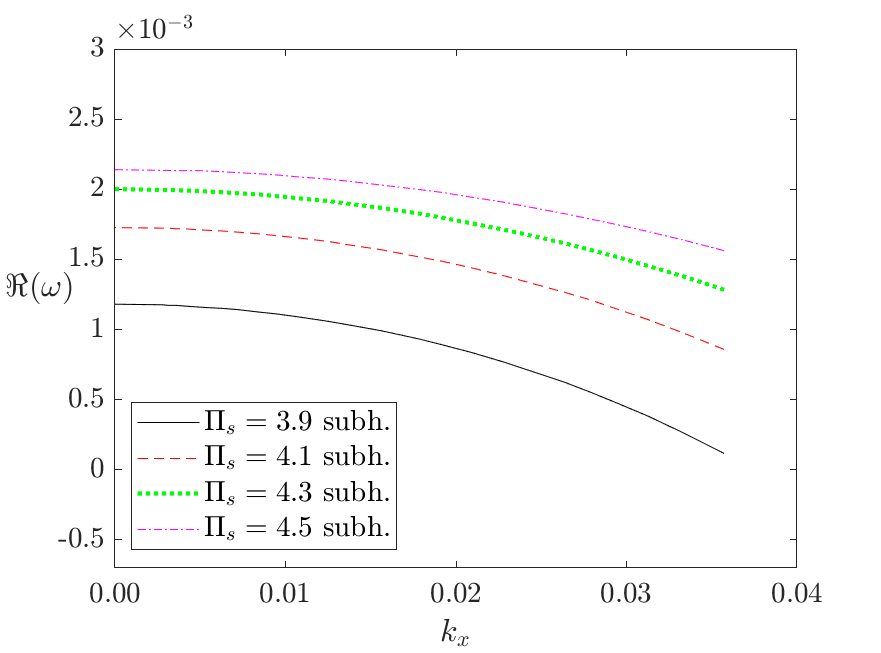}
   \caption{ }
	\label{subfig:subh22_gr}
 \end{subfigure}
 \vfill
 \vspace{15pt}
   \begin{subfigure}{0.5\textwidth}
\centering
	\includegraphics[width=1\textwidth]{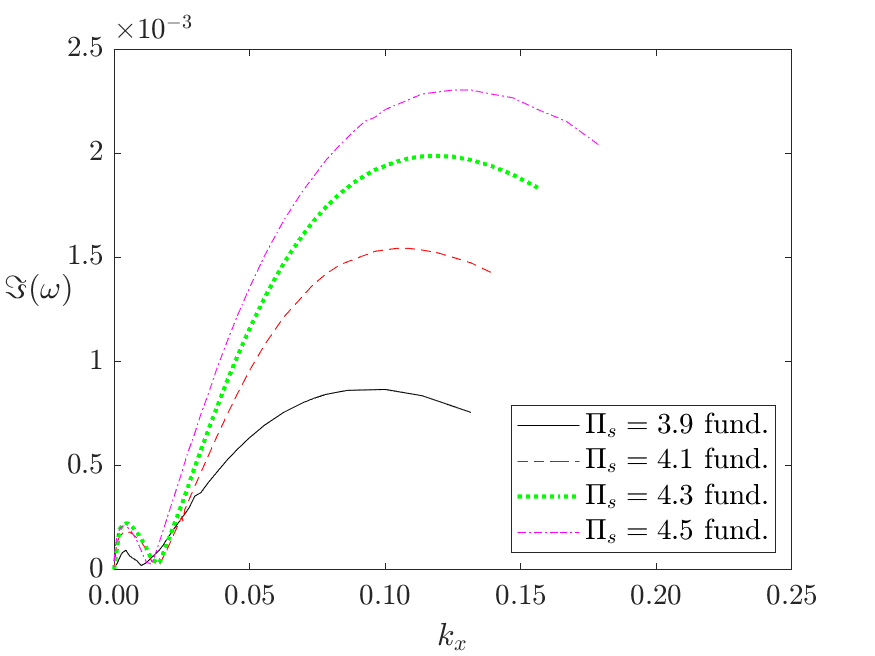}
   	\caption{}
	\label{subfig:fund22_osc}
 \end{subfigure}
    \begin{subfigure}{0.5\textwidth}
	\includegraphics[width=1\textwidth]{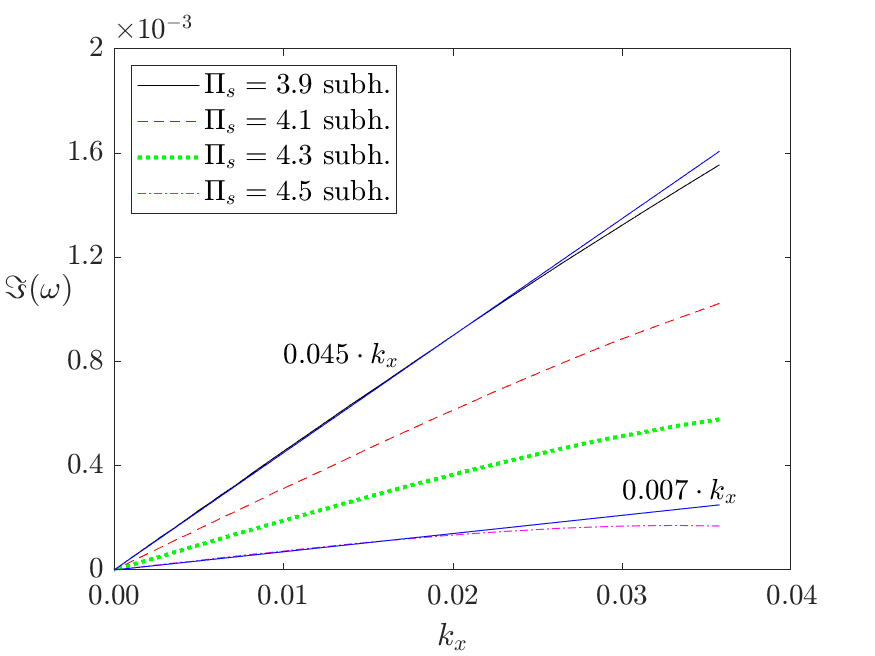}
   	\caption{}
	\label{subfig:subh22_osc}
 \end{subfigure}
 
	\caption{ Comparison between fundamental and subharmonic modes of katabatic speaker-wire vortices at slope angle $\alpha=22^{\circ}$   for different values of $\Pi_s$ as a function of the longitudinal wave number $k_x$  : (a)-(b) Growth rates $\Re(\omega)$;  (c)-(d) oscillation frequency $\Im(\omega)$. Fundamental modes are shown on the left(a,c), and subharmonic modes are shown on the right(b,d). The linear fit for the dispersion relation of the subharmonic mode is also shown in (d).  }
\end{figure}

The  growth rates of the most unstable fundamental modes as a function of streamwise wavenumbers  $k_x$ within the range $[0,0.2]$ are shown 
for four different values of normalized surface heat flux $\Pi_s =3.9,4.1,4.3,4.5$
in figure \ref{subfig:fund22_gr}. Again, a higher $\Pi_s$ value than at the lower slope angles is needed to trigger and maintain the primary roll instability for the base vortices.  The  growth rates of the strongest subharmonic modes for streamwise wavenumbers  $k_x$ within  $[0,0.04]$ are shown  in figure \ref{subfig:subh22_gr}. It can be seen that  a higher  value of $\Pi_s$   strengthens all instability modes at any given wave number. By comparing the growth rate plots shown in Figs. \ref{subfig:fund22_gr} and \ref{subfig:subh22_gr}  side by side,
 We observe that  the fundamental modes and subharamonic modes are of a similar strength across all wave numbers. The main distinction between the  two mode types is the fact that the strongest fundamental mode are 3-D and achieve the highest growth at an optimal non-zero wave number $k_x$ between 0.02 for $\Pi_s=3.9$  and 0.11 for $\Pi_s=4.5$, displaying a clear trend of increasing streamwise wavelength with decreasing $\Pi_s$ value. On the other hand, the subharmonic modes become monotically stronger with decreasing wave number such that the most dominant mode is purely 2-D at $k_x=0$, same as in the previously studied slope angles. 

All the  fundamental modes with non-zero streamwise wavenumbers studied in this case are  oscillatory. However, the two-dimensional mode with $k_x=0$ is stationary.  As shown in figure \ref{subfig:fund22_osc}, within the very small wavenumber range $0<k_x<0.01$, the frequencies for modes with different $\Pi_s$ values  attain a local maximum at $k_x\approx 0.005$. In the  wavenumber range $0.01<k_x<0.08$,  the fundamental mode frequencies   increase with growing wavenumber. However, at higher wavenumbers $k_x>0.1$, this trend no longer holds, and the frequencies even slightly decrease with increasing $k_x$. In contrast, figure \ref{subfig:subh22_osc} indicates that the frequency of subharmonic modes is a monotonically growing function of their wavenumber $k_x$. All 2-D subharmonic modes ($k_x=0$) are stationary, and  for the low wavenumber range, the   frequencies for all four $\Pi_s$ values can be described via a linear dispersion relation given by $\Im(\omega)=\eta\cdot k_x$. In contrast to the previous case at smaller slope angle, however, the group velocity  $\eta$ decreases with growing $\Pi_s$   from $\eta=0.045$ at $\Pi_s=3.9$  to $\eta=0.007$ for $\Pi_s=4.5$.  At higher wavenumbers beyond $k_x=0.03$,  the oscillation frequencies of the subharmonic modes start to grow slower than the previous linear rate with increasing wavenumbers.  

From these computed properties of fundamental and subharmonic secondary instabilities, we can derive helpful inferences about the dynamics of vortices at the steep slope angle of $\alpha=22^{\circ}$. 
Since the strongest subharmonic mode dominate at lower wave numbers,  we can expect long-wave reconnections  or pure mergers between neighboring vortex pairs involving four vortices. The fundamental modes which are stronger at shorter wavelengths will bend each individual vortex parallel to its adjacent neighbors. In the full nonlinear vortex evolution however, it turns out that the the subharmonic modes appear to have a stronger impact on the vortex dynamics than predicted from linear stability analysis, as vortex mergers and reconnections seem to be the dominant feature of such flows with little signature of the wavy bending signature of the fundamental modes. This is very similar to the dynamics at higher slope angles as described below and shown in the animation contained in supplementary movie 3.

\subsubsection{ Very steep slopes $\alpha=30^{\circ}$}

\begin{figure}

 \vspace{15pt}
 \centering
   \begin{subfigure}{0.7\textwidth}
		
	\includegraphics[width=0.99\textwidth]{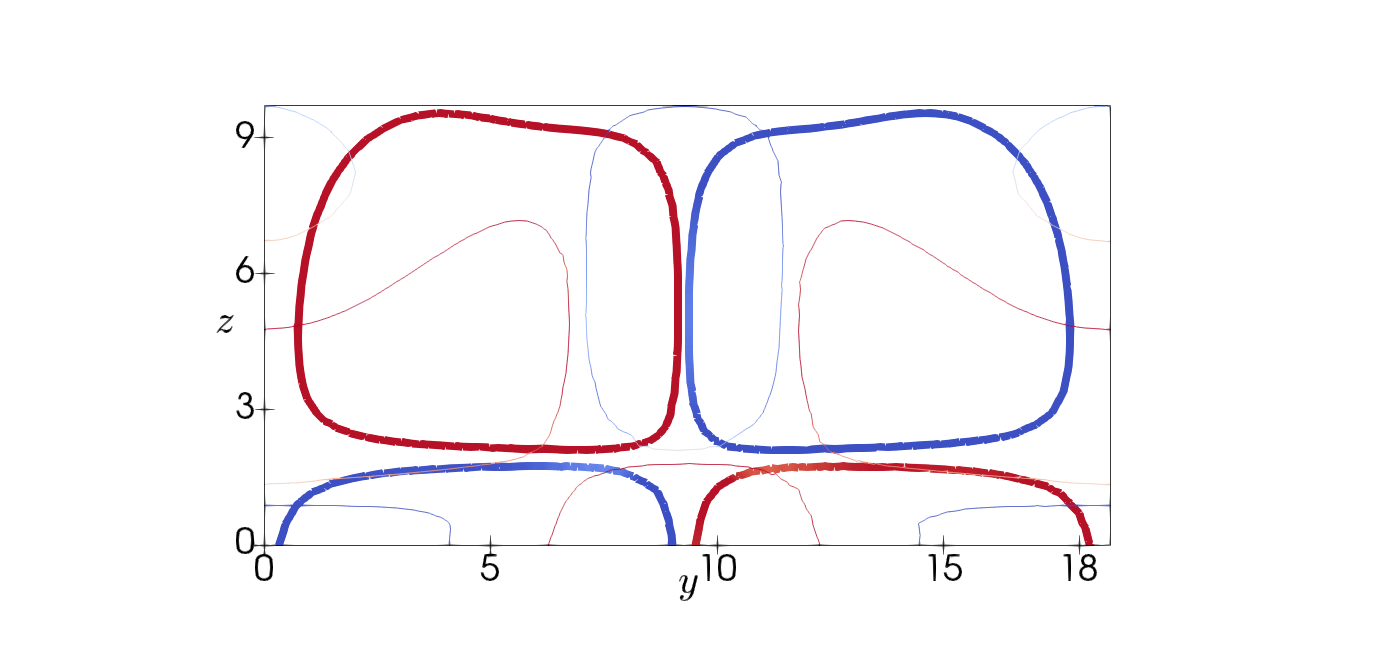}
   	\caption{}
	\label{subfig:fund30_contour2d}
 \end{subfigure}
  
      \begin{subfigure}{0.65\textwidth}
      	\centering
	\includegraphics[width=0.99\textwidth]{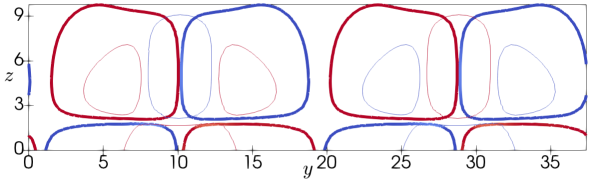}
   	\caption{}
	\label{subfig:subh30_contour2d}
 \end{subfigure}
 
	\caption{ Visualization  of streamwise vorticity magnitude for the fundamental mode on cross-slope $yz$ vertical plane  in katabatic  flow  at   $\alpha=30^{\circ}$ and $\Pi_s=6.1$: (a) fundamental mode with $k_x=0.06$; (b) subharmonic mode with $k_x=0.03$. Thick contour lines represent the base flow, and thin lines are from the instability mode.  Contours are drawn at 8\% of maximal magnitude in both cases. 
 } 
\end{figure}

The  subharmonic and fundamental  instabilities of speaker-wire vortices at the steepest slope  in this study with $\alpha=30^{\circ}$ are  visualized together with the base vortices via the contour plot for the streamwise vorticity $\omega_x$ on the transverse plane  in Figs. \ref{subfig:fund30_contour2d} and \ref{subfig:subh30_contour2d}.  
  When compared to the  previous cases at smaller angles as shown in Fig. \ref{fig:fund03_contour2d}, Figs. \ref{subfig:fund12_contour2d}-\ref{subfig:subh12_contour2d} and Figs. \ref{subfig:fund22_contour2d}-\ref{subfig:subh22_contour2d}, it can be seen that the  base flow vortices continue to shrink along the transverse direction and grow in the vertical  with increasing slope angle. 
As a consequence,  the transverse wavelengths  of the fundamental and subharmonic modes are also clearly less than that of the modes at lower slope angles.

\begin{figure}
 \begin{subfigure}{0.5\textwidth}

	\centering
	\includegraphics[width=1\textwidth]{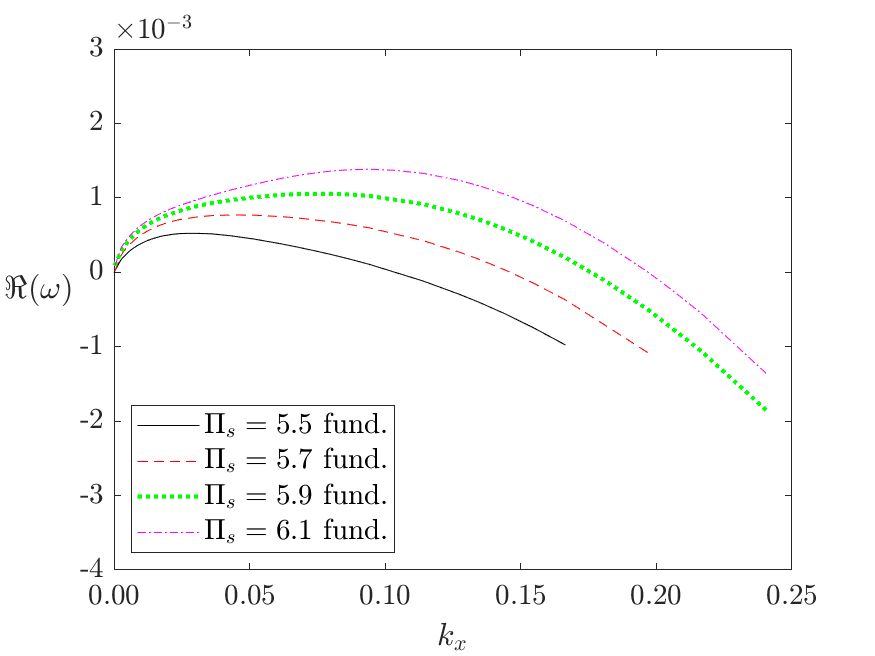}
   \caption{ }
	\label{subfig:fund30_gr}
 \end{subfigure}
  \begin{subfigure}{0.5\textwidth}
	\centering
	\includegraphics[width=1\textwidth]{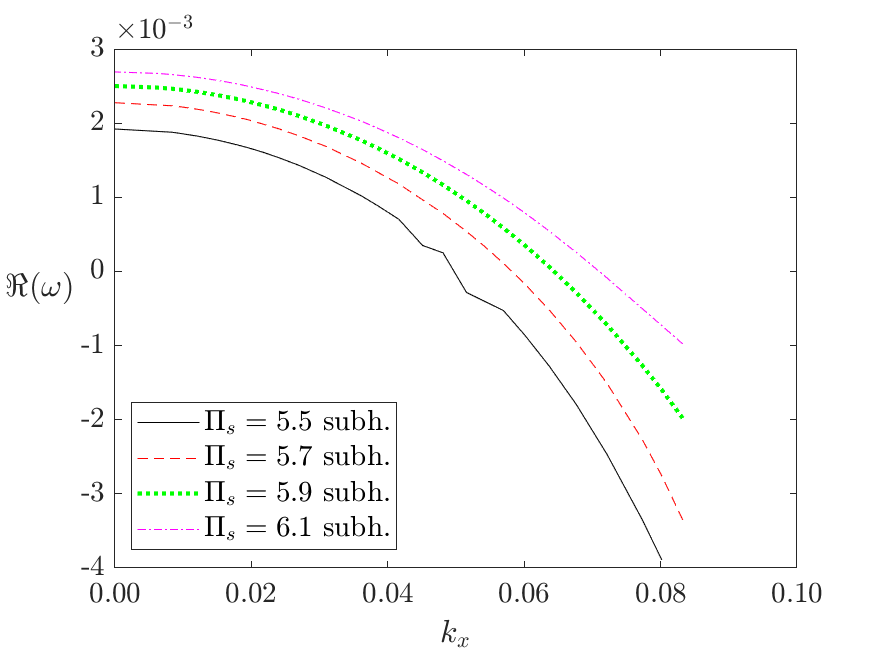}
   \caption{ }
	\label{subfig:subh30_gr}
 \end{subfigure}
 
 \vspace{15pt}
   \begin{subfigure}{0.5\textwidth}
\centering
	\includegraphics[width=1\textwidth]{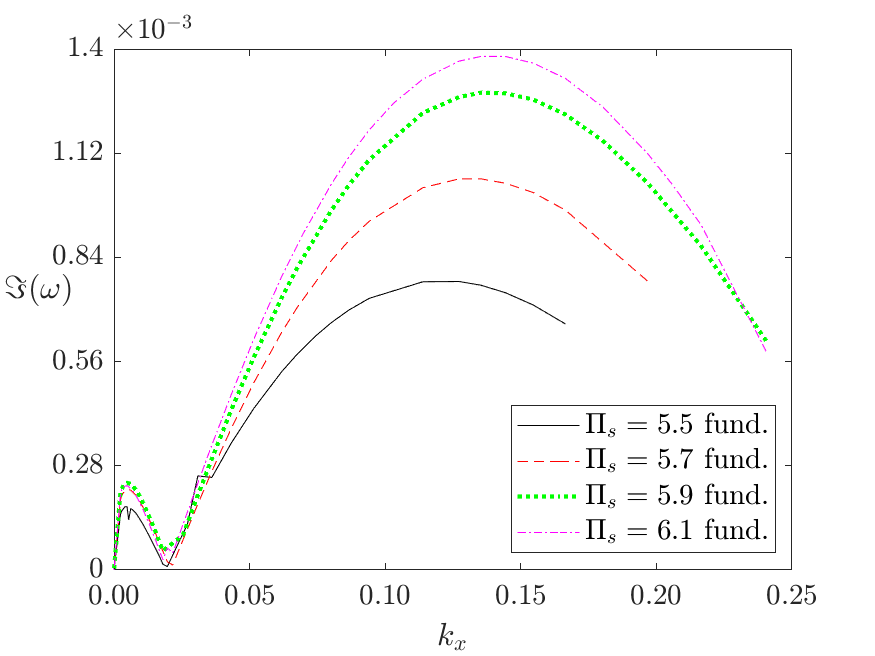}
   	\caption{}
	\label{subfig:fund30_osc}
 \end{subfigure}
    \begin{subfigure}{0.5\textwidth}
	\includegraphics[width=1\textwidth]{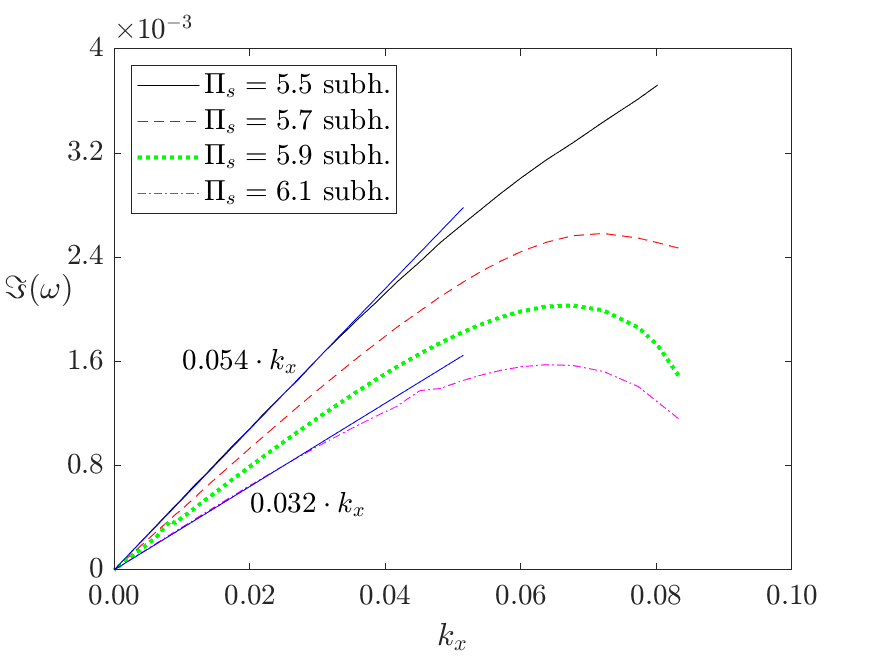}
   	\caption{}
	\label{subfig:subh30_osc}
 \end{subfigure}
 
	\caption{ Comparison between fundamental and subharmonic modes of katabatic speaker-wire vortices at slope angle $\alpha=30^{\circ}$   for different values of $\Pi_s$ as a function of the longitudinal wave number $k_x$  : (a)-(b) Growth rates $\Re(\omega)$;  (c)-(d) oscillation frequency $\Im(\omega)$. Fundamental modes are shown on the left(a,c), and subharmonic modes are shown on the right(b,d). The linear fit for the dispersion relation of the subharmonic mode is also shown in  (d). }
\end{figure}

The  growth rates of the most unstable fundamental modes as a function of streamwise wavenumbers  $k_x$ within the range $[0,0.25]$ for four different values of normalized surface heat flux $\Pi_s =5.5,5.7,5.9,6.1$  are shown  in figure \ref{subfig:fund30_gr}. The  growth rates of the strongest subharmonic modes for streamwise wavenumbers  $k_x$ within  $[0,0.1]$ are shown  in figure \ref{subfig:subh30_gr}. Similar like in the previous cases, the strongest subharmonic instabilities have larger streamwise wavelengths than their fundamental counterparts.
 For sufficiently small wavenumbers, i.e. $0<k_x<0.06$, it can be seen that the subharmonic modes are clearly stronger than the fundamental ones, with the distinctively dominating 2-D subharmonic mode at $k_x=0$. However, since the growth rate of subharmonic modes decays rapidly with growing wavenumber whereas the growth rates of fundamental modes are far less sensitive to wavenumber variation,   at sufficiently large wavenumbers $k_x>0.1$,  only the fundamental modes have positive growth rates.

All  fundamental modes with non-zero longitudinal wavenumbers $k_x>0$ are oscillatory; their frequencies  are doubly-parabolic functions of $k_x$ with one  local maximum in the low-wavenumber range at  $k_x\approx 0.01$ and a global maximum at $k_x\approx 0.15$, as shown in figure \ref{subfig:fund30_osc}. On the other hand, figure \ref{subfig:subh30_osc} shows that the frequencies of subharmonic modes have a less complicated relationship to their wavenumbers. All 2-D subharmonic modes($k_x=0$) are stationary, and  for wavenumber values less than $k_x=0.04$, the  normalised frequencies for all four $\Pi_s$ values seem to fit a linear dispersion relation given by $\Im(\omega)=\eta\cdot k_x$, where the group velocity is $\eta=0.054$ for $\Pi_s=5.5$ and decreases  with growing $\Pi_s$ to $\eta=0.032$ at $\Pi_s=6.1$.  However, at larger wavenumbers $k_x>0.04$, this linear relation no longer holds true.  For $\Pi_s=5.5$, the frequency increases at a smaller rate with growing  $k_x$ whereas for the cases with larger $\Pi_s$ values, the frequency starts to decrease with growing wavenumber when $k_x>0.06$.

From these  properties of fundamental and subharmonic vortex instabilities at the very steep slope, we may infer that 2-D and long-wave subharmonic modes are the dominant driver of vortex dynamics. They would manifest themselves as pure mergers or long-wave reconnections between neighboring vortex pairs. Even though the linear stability results may suggest that the fundamental modes to be visible at shorter wavelengths, in reality, they are overshadowed by the effect of the far stronger long-wave subharmonic modes. 
This   can be seen  from animations obtained from direct numerical integration  of the full nonlinear governing equations  in which the dominating dynamics are vortex mergers and long-wave instabilities,  as shown in the animation available as supplementary movie 3.

\subsection{ Vortex dynamics due to  fundamental and subharmonic modes }

The role of fundamental instabilities (i.e. modes which have the same transverse spatial period  as the base flow) on the dynamics of an array of longitudinal rolls  has been extensively documented in  previous studies. 
As an example, they have also been identified as instabilities of  Rayleigh-Benard convection rolls  which aim at distorting the structure and spacing of the rolls to bring them closer to the optimal wavelength \cite{clever1974transition, pierrehumbert1982vortexinstab}.
Other well-known representatives for such modes include the  Crow instabilities which are responsible for the bending and reconnection of  a  vortex pair suspended in quiescent air \cite{crow1970stability}. 

\begin{figure}

 \vspace{15pt}
   \begin{subfigure}{0.48\textwidth}
		\centering
	\includegraphics[width=0.99\textwidth]{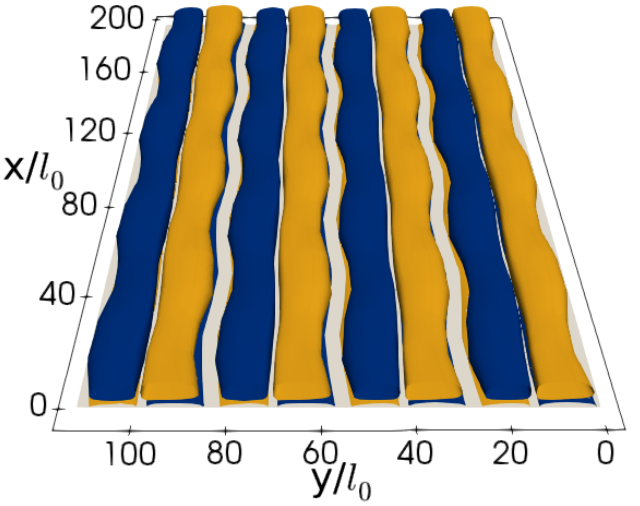}
   	\caption{}
	\label{subfig:fund12_contour3d}
 \end{subfigure}
  \hfill
      \begin{subfigure}{0.47\textwidth}
      	\centering
	\includegraphics[width=0.7\textwidth]{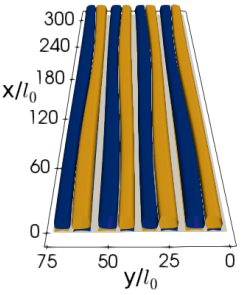}
   	\caption{}
	\label{subfig:subh30_contour3d}
 \end{subfigure}
 
	\caption{ Distortion of longitudinal speaker-wire vortices in katabatic  flows due to instabilities: Effect of (a) Fundamental mode at $\alpha=12^{\circ}, \Pi_s=2.9$ and (b) subharmonic mode at  $\alpha=30^{\circ}, \Pi_s=6.1$. 
 Visualization of streamwise vorticity contours at 8\% of maximal magnitude for base vortices added with the corresponding instability modes.  } 
\end{figure}

The effect of a fundamental instability on vortices at a slope angle of $12^{\circ}$, which is representative for fundamental modes at other angles as well, is visualised in Fig. \ref{subfig:fund12_contour3d}. As can be seen, this 3-D fundamental mode  causes  sinusoidal bending and distortion of each speaker-wire vortex.  
It is worth noticing that the fundamental modes for speaker-wire vortices are anti-symmetric, i.e. they can only bend both vortices within a pair along the same direction, as shown in Fig. \ref{subfig:fund12_contour3d}.

Like their fundamental counterparts, 2-D and 3-D subharmonic vortex instabilities which have twice the transverse wavelength as the base flow also  play a prominent role in shaping the vortex dynamics, such as in co-rotating Stuart vortex arrays  \cite{pierrehumbert1982vortexinstab} or as secondary instabilities in a shear layer by \cite{corcos1984mixing}, where they are shown to be responsible for the merging of neighboring vortices. As visualised in  Fig. \ref{subfig:subh30_contour3d}, the 3-D secondary subharmonic modes of the speaker-wire vortices in katabatic slope flows  bend adjacent speaker-wire vortices in opposite directions  as to facilitate their reconnection in the 3-D case and merger for the 2-D mode. In a recent investigation, the presence of long-wave vortex reconnections associated with the low-frequency content in flow spectra has been observed in the DNS of steep katabatic slope flows with angle $\alpha>20^{\circ}$ \citep{henao2023investigation}.

We observe from Fig. \ref{subfig:fund12_contour3d} and \ref{subfig:subh30_contour3d} that  both the fundamental  and subharmonic modes are anti-symmetric within a single counter-rotating pair, i.e. they bend the two sister rolls of the same pair along the same direction, thus preventing a vortex reconnection or merger of neighboring vortices within the same pair in contrast to what is observed for a single vortex pair with Crow instability \citep{crow1970stability}. This means that a single vortex pair (i.e. a speaker-wire vortex) can remain in its basic pair structure even after the initial onset of instabilities, thus justifying their designation as a coherent vortex structure. Similar vortex structures which remains stable and coherent over long wavelengths have been observed in the \textit{Langmuir vortices} on the surface of seas and oceans as described in \cite{craik_leibovich1976langmuir}.\\ 

 As outlined in the previous discussions, the properties of secondary modes for steady speaker-wire vortices are heavily influenced by the size of the slope angle $\alpha$. This in turn translates into different vortex dynamics at different slope angles.  For the four different slope angles  $\alpha=3^{\circ},12^{\circ},22^{\circ},30^{\circ}$ studied here, our analysis has shown that with increasing values of $\alpha$,  the subharmonic modes become more dominant relative to  their fundamental counterparts. At shallow slopes with $\alpha=3^{\circ}$, only   fundamental modes are supported, whereas at  steep slopes with $\alpha>30^{\circ}$, the subharmonic modes have the largest growth rates and  are clearly stronger than their fundamental counterparts in the  low wavenumber range.  
 When $\alpha$ takes on an intermediate value such as between $10^{\circ}$ and $20^{\circ}$, both fundamental and subharmonic modes have similar growth rates   at  stratification parameter $\Pi_s$ not far above the linear  stability threshold. To track the further evolution of the vortex dynamics beyond the initial destabilization by the instability modes, we have carried out a series 3-D numerical simulations based on the full nonlinear set of equations (\ref{eqnslopemom})-((\ref{eqnslopecont}), initializing the flow field with an array of at least 4 base vortex pairs upon which a fundamental or subharmonic mode with $10\%$  of the base flow strength has been added. The simulation domain and boundary conditions for  cross-slope $yz$ plane are the same as described in section \ref{sec_sim2d}, while 32 Fourier modes have been used to resolve the along-slope $x$ direction, which is sufficient for our purpose to study the higher order vortex dynamics before the onset of full turbulence. The animated simulation in supplementary movie 1 shows the initial growth of the fundamental mode in longitudinal rolls at the slope angle $\alpha=3^{\circ}$, which later leads to novel structures on top of the original base vortices. Supplementary movie 2 shows the initial vortex dynamics at $\alpha=12^{\circ}$ due to to the anti-symmetric fundamental mode, which later makes way for a long-wave subharmonic mode that merges two adjacent vortex pairs.
 The animation shown in supplementary movie 3 illustrates how the subharmonic mode at $\alpha=30^{\circ}$ causes vortex reconnections and mergers between rolls from two adjacent pairs, without any apparent  signature   from a short-wave fundamental mode such as sinuosoidal bending of vortices. 
 
\subsection{Comparison of secondary vortex modes under katabatic and anabatic conditions} 

In our earlier work \citep{xiao2020anabatic}, we have established that in anabatic slope flows, the stationary roll mode is only the dominant primary instability at slope angle values less than 10 degrees  whereas at steeper slopes $\alpha >10^{\circ}$, the 2-D travelling wave mode would replace it as the stronger instability. On the other hand,   under katabatic slope flow conditions, the stationary longitudinal  rolls can emerge naturally as a result of the strongest instability mode over a wide range of slope angles up to 70 degrees. Thus, a comparison between the dynamics  of longitudinal rolls under these two conditions is only possible for a narrow range of shallow slopes less than $10^{\circ}$. In this work, the slope angle $\alpha=3^{\circ}$ is chosen as the common reference point to contrast anabatic and katabatic slope flows, which is a realistic slope angle that can be observed in actual terrains. For further comparison purposes, we also introduce vortices formed under katabatic conditions at the steep slope angle $\alpha=30^{\circ}$.   
 
\begin{figure}
\centering
 \begin{subfigure}{0.5\textwidth}
	\centering
	\includegraphics[width=1\textwidth]{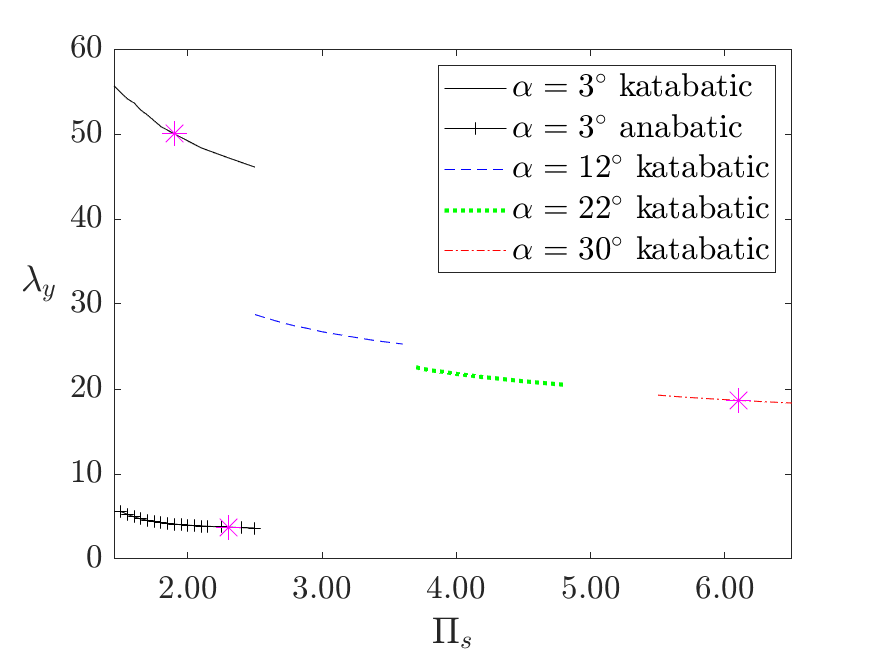}
   \caption{ }
	\label{subfig:slope3_comparekatanal}
 \end{subfigure}
   \begin{subfigure}{0.75\textwidth}
	\centering
	\includegraphics[width=1\textwidth]{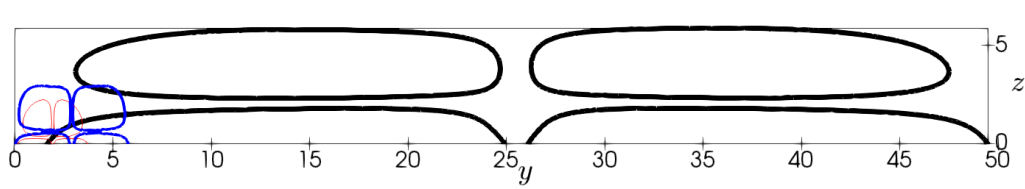}
   \caption{ }
	\label{subfig:slope3_comparekatana2d}
 \end{subfigure}
 	\caption{Comparison of the most dominant primary  instabilities of Prandtl slope flows under  katabatic  and anabatic conditions leading to longitudinal rolls, at different slope angles $\alpha$  and as a function of $\Pi_s$: (a) transverse wavelength of the most dominant mode, with magenta asterisks indicating cases for which steady vortices have been obtained through numerical simulations; 
	(b)  transverse cross-sections  of longitudinal speaker-wire   vortices arising from 1-D Prandtl flow profile for katabatic  and anabatic slope flows at different angles for the three cases marked by asterisks in (a). Reference length scale is chosen to be the value at $\alpha=3^{\circ}$ as given in equation (\ref{eqnlscale}). Streamwise vorticity contour at 12\% of respective maximal magnitude is visualized. In (b), thick black lines represent katabatic flow at $\alpha=3^{\circ}, \Pi_s=1.9$; medium blue lines represent katabatic flow at $\alpha=30^{\circ}, \Pi_s=6.1$, and thin red lines represent anabatic flow with $\alpha=3^{\circ}, \Pi_s=1.9$;  }
\end{figure}

To illustrate the most significant differences between katabatic and anabatic slope flow conditions, we show the optimal transverse wavelengths of the primary roll instability for the one-dimensional Prandtl flow profile at different slope angles as a function of $\Pi_s$ in Fig. ~\ref{subfig:slope3_comparekatanal}. The aforementioned trend of decreasing vortex spacing with increasing slope angle for katabatic  flows is clearly evident, and for both katabatic as well as anabatic conditions, the optimal vortex width decreases with increasing $\Pi_s$ value.

At the shallow slope with angle $\alpha=3^{\circ}$ where steady vortices can form under both katabatic and anabatic  conditions, we observe from Fig. ~\ref{subfig:slope3_comparekatanal} that for the same value of $\Pi_s=1.9$ but with opposite  signs of surface heat flux, the   transverse wavelength $\lambda_y$ of the most dominant vortex  in the katabatic case is an order of magnitude larger than under anabatic conditions. This observation is visualized in Fig.  \ref{subfig:slope3_comparekatana2d}, which displays the streamwise vorticity contours of both anabatic and katabatic vortices at $\alpha=3^{\circ}$, and it is obvious that the vortices formed under katabatic conditions are multiple times wider and taller. Fig.  \ref{subfig:slope3_comparekatana2d} also shows that only at a far steeper slope angle of $\alpha=30^{\circ}$ do the katabatic vortices assume a similar size as their anabatic  counterparts at $\alpha=3^{\circ}$.

\begin{figure}
 \begin{subfigure}{0.5\textwidth}

	\centering
	\includegraphics[width=1\textwidth]{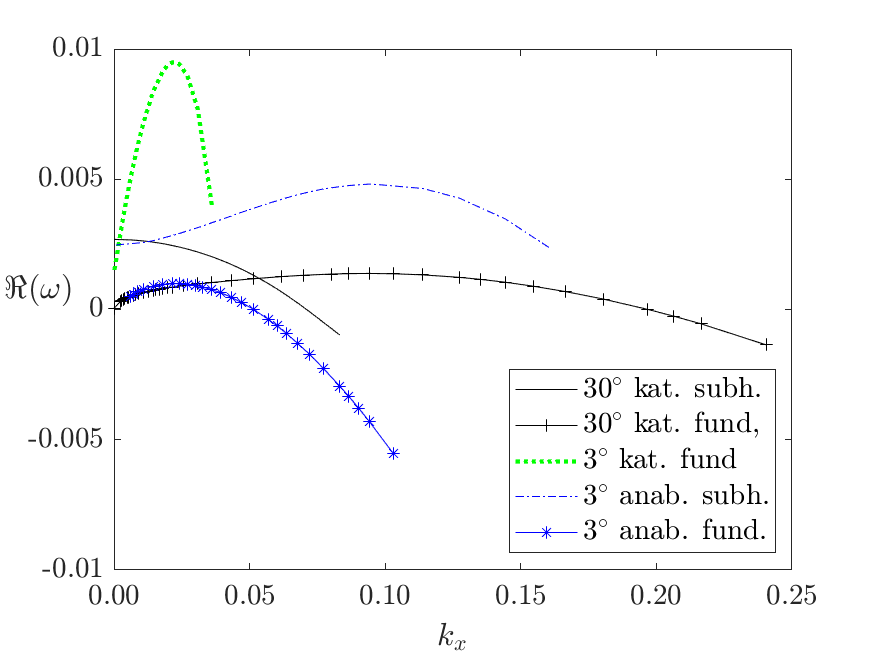}
   \caption{ }
	\label{subfig:slope3_companakat_gr}
 \end{subfigure}
  \begin{subfigure}{0.5\textwidth}
	\centering
	\includegraphics[width=1\textwidth]{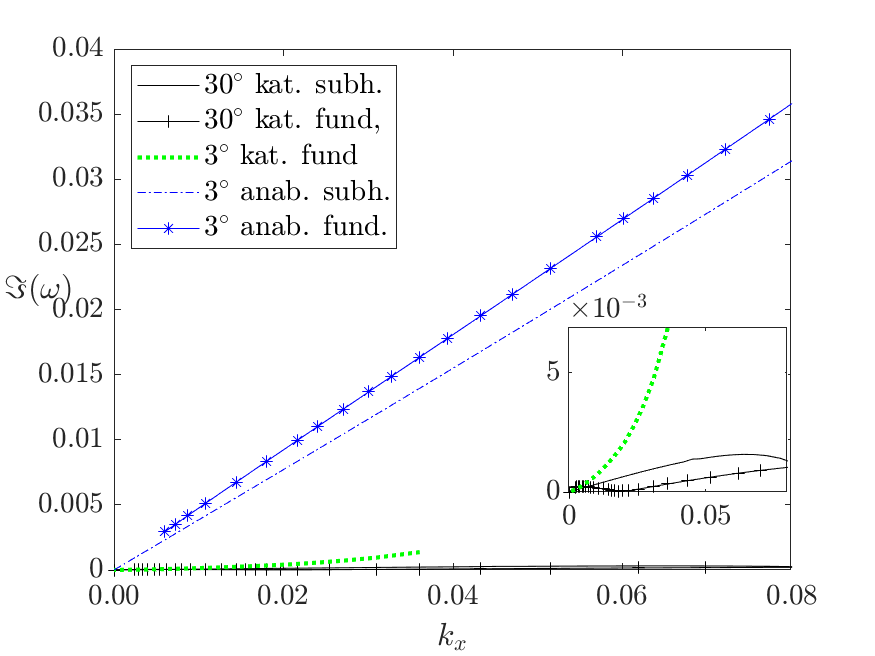}
   \caption{ }
	\label{subfig:slope3_companakat_osc}
 \end{subfigure}
 
	\caption{Comparison of  instability modes for the longitudinal base vortices  under  katabatic  and anabatic conditions in Prandtl slope flows, at different slope angles $\alpha$  and as a function of streamwise wave number $k_x$: (a) growth rate; (b) oscillation frequency (inset shows detailed view for katabatic conditions). The values of $\Pi_s$ are the same for the corresponding vortices as in the previous figure.  }
\end{figure}

The growth rates of secondary vortex instabilities under katabatic and anabatic conditions are displayed in Fig. ~\ref{subfig:slope3_companakat_gr}. 
For the same slope angle of $3^{\circ}$ and  stratification parameter  $\Pi_s=1.9$, there are no instability modes for the base vortices at their most preferred width under anabatic conditions, whereas the katabatic configuration is subject to a fundamental secondary instability.  Hence,  a higher value of $\Pi_s=2.3$ is chosen for the anabatic slope flow in order to compare its vortex instability against the katabatic conditions at $\Pi_s=1.9$.      
As described previously, there exists no subharmonic mode for katabatic vortices  at $\alpha=3^{\circ}$, whereas the anabatic vortices are susceptible to both fundamental and subharmonic instabilities. Fig. ~\ref{subfig:slope3_companakat_gr} shows that even though the katabatic configuration has a lower $\Pi_s$ value, its secondary instability exhibits an up to four times  larger growth rate than the strongest modes for anabatic vortices. Another key distinction is that the most dominant mode in the katabatic case is clearly 3-D since its highest growth rate at the non-zero optimal streamwise wavenumber is clearly larger than the growth rate of the 2-D mode with zero wavenumber. There is no such behavior in the anabatic case where for both fundamental and subharmonic modes, the largest growth rates are attained at the lowest wavenumbers, and  the growth rates remain relatively constant  for low wavenumbers. It is interesting to notice that for katabatic vortices at  $\alpha=30^{\circ}$ which are of a similar size as the anabatic vortices at the shallow $\alpha=3^{\circ}$, the growth rates of the fundamental and subharmonic modes at the low wavenumber range are comparable to those under anabatic conditions, which indicates that the size and shape of the base vortices rather than the slope angle directly exerts a major influence on secondary mode growth for long-wave modes. 

The oscillation frequencies of secondary vortex instabilities under katabatic and anabatic conditions are plotted in figure~\ref{subfig:slope3_companakat_osc}. It is evident that the secondary modes under anabatic conditions have clearly higher frequencies at all wavenumbers than  those of modes for katabatic vortices. A major distinction is that while the frequencies of both subharmonic and fundamental modes in the anabatic case fit a linear dispersion relation over the entire range of streamwise wavenumber $0<k_x<0.08$, no such simple dependency exists for the frequencies of secondary modes under katabatic conditions which can increase at different rates or even decrease with growing $k_x$.

\section{Conclusions}
 
We carried out a bi-global linear stability analysis to gain insights into the dynamics of longitudinal rolls, a.k.a. speaker-wire vortices, that emerge due to a primary instability from the  one-dimensional katabatic slope flows over a range of slope angles. The katabatic conditions are the analogue to the anabatic slope flow  conditions, whose effects on vortex dynamics are described in our prior work \citep{xiao2022speaker}. 

Our base flow configuration under katabatic conditions is uniquely different than other well-known counter-rotating vortex pairs \citep{crow1970stability,billant2010zigzag1,hattori2021modal,busse1979instabilities,clever1974transition}. The katabatic Prandtl slope flow includes an independent background stratification that is at an angle to the cooled solid slope surface.   Another unique feature of our configuration is the fact that the stationary longitudinal rolls serving as base flow have three non-zero velocity components even though the flow field is still 2-D, i.e. it only varies along the vertical and transverse dimensions. As a result, it is not too surprising that the instability dynamics of speaker-wire vortices investigated in the current work are  distinct from the hitherto known vortex-pair instabilities.

Similar to our earlier investigation of speaker-wire vortices under anabatic conditions,  we have established that  the counter-rotating vortex pair form a coherent stable unit (i.e. a speaker-wire vortex) that can only lose their individual rolls to mergers or reconnections in the presence of another speaker-wire vortex. Our results for secondary instabilities of speaker-wire vortices have shown that the slope angle $\alpha$ as well as the dimensionless stratification perturbation parameter $\Pi_s$, which can be interpreted as a normalized surface heat flux or the strength of the surface thermal forcing relative to the ambient stratification,  play a major role in determining which modes are the most significant in destabilizing vortex rolls formed in katabatic slope flows. The major distinction between vortex instabilities in the  anabatic and katabatic configurations is the fact  that in the latter case, stationary vortices can emerge from the 1D Prandtl base flow profile for all slope angles less than around $70^{\circ}$, whereas this is restricted only to non-steep slopes less than $10^{\circ}$ for anabatic slope flows.  

 For slope angles in the range of $3^{\circ}-30^{\circ}$ at which the 1D katabatic Prandtl slope flow profile is naturally susceptible to the stationary roll mode as primary instability, the most dominant vortex instability shifts from the fundamental mode towards the  subharmonic mode with longer steamwise wavelengths with increasing slope steepness. For a shallow slope with $\alpha=3^{\circ}$, the only unstable mode is the fundamental instability, no matter how strong the imposed surface heat flux is. This contrasts with vortices in anabatic slope flows at the same angle for which the subharmonic instability is dominant. For larger slope angles, the subharmonic mode begins to emerge as well. At $\alpha=12^{\circ}$, subharmonic mode is still clearly weaker than its fundamental counterpart except at the smallest wave numbers. For even steeper slopes with $\alpha = 22^{\circ}$, both subharmonic modes and fundamental modes are of comparable strength over a broad range of wave numbers. At the steepest slope angle $\alpha=30^{\circ}$  investigated in this work, the subharmonic mode is clearly more dominant than its fundamental counterpart except at the largest wavenumbers. The main difference between the two mode types is that the fundamental  instability is a 3-D oscillatory mode whose growth rate  at the optimal nonzero streamwise wavenumber is many times larger than the growth rate of the 2-D zero wave number mode. In contrast, the subharmonic mode attains its strongest growth at zero streamwise wavenumber where its oscillation frequency is also zero, hence it is a 2-D stationary mode.  This mode exhibits its two-dimensional nature by  directly merging  two entire rolls from two adjacent but separate speaker-wire vortices without any bending in the streamwise direction.

The 3-D fundamental mode, with a transverse wavelength equivalent to the vortex separation of the base speaker-wire vortices, is anti-symmetric and bends all  rolls in all speaker-wire vortices along the same direction, thus the distances between  them remain the same as in the original base configuration before the onset of instability. Due to the absence of any symmetric instabilities, the fundamental mode cannot bend the two vortices within a pair towards each other. However, after sufficiently long time, the long-wave or two-dimensional subharmonic instability   will eventually manifest itself to  move two neighboring speaker-wire vortices from different pairs towards each other,  resulting in the merger between two adjacent rolls which are not from the same speaker-wire vortex as in the base configuration as described above.
 
 Our results  demonstrate that the only possible  vortex merger or reconnection dynamics under Prandtl's katbatic slope flow model is triggered by a long-wave subharmonic mode which requires  four vortex rolls in two speaker-wire vortices. In contrast, one single speaker-wire vortex is able to maintain its two-roll structure even after the onset of the fundamental  instabilities. The dependence of vortex dynamics on the slope angle is an extension of our earlier study for vortices in anabatic slope flows which focused on the effect of vortex width and separation \citep{xiao2022speaker}.  Hence, the  vortex dynamics under katabatic conditions investigated in this work are expected to contribute toward a better understanding of turbulent transition in stably stratified boundary layers on non-flat surfaces.

\textbf{Supplementary movies.} 3 Supplementary movies have been attached.\\

\textbf{Acknowledgments:}
This material is based upon work supported by the National Science Foundation under Grant No. (1936445). Research was sponsored in part by the University of Pittsburgh, Center for Research Computing through the computing resources provided.

\textbf{Declaration of Interests}: The authors report no conflict of interest.

\bibliographystyle{jfm}
\bibliography{stability}

\end{document}